%% file: revised_ms.tex
\documentclass[aps,prb,article,twocolumn,showpacs,preprintnumbers,amsmath,amssymb,superscriptaddress]{revtex4-1}
\date{\today}
\usepackage{epsfig}
\usepackage{epstopdf}
\usepackage{subfigure}
\usepackage{graphicx}
\usepackage{dcolumn}
\usepackage{bm}
\usepackage[colorlinks,linkcolor=blue,hyperindex,CJKbookmarks]{hyperref}
\usepackage{float}
\usepackage{comment}
\usepackage{xcolor}
\usepackage[textsize=footnotesize,textwidth=0.5in]{todonotes}
\usepackage{etoolbox}

\begin{document}

\title{Influence of multiplet structure on photoemission spectra of
spin-orbit driven Mott insulators: application to \boldmath${\rm Sr_2IrO_4}$}

\author{Ekaterina M. P{\"a}rschke}
 \email[]{ekaterina.paerschke@gmail.com}
 \affiliation{IFW Dresden, Helmholtzstr. 20, 01069 Dresden, Germany}
 \affiliation{Department of Physics, University of Alabama at Birmingham, Birmingham, Alabama 35294, USA}

\author{Rajyavardhan Ray}
 \email[]{r.ray@ifw-dresden.de}
 \affiliation{IFW Dresden, Helmholtzstr. 20, 01069 Dresden, Germany}
 \affiliation{Dresden Center for Computational Material Science (DCMS), TU Dresden, 01062 Dresden, Germany}

\date{\today}
\begin{abstract}
Most of the low-energy effective descriptions of spin-orbit driven Mott
insulators consider spin-orbit coupling (SOC) as a second-order
perturbation to electron-electron interactions. However, when SOC is
comparable to anisotropic Hund's coupling, such as in Ir, the validity
of this formally weak-SOC approach is not a priori known. Depending
on the relative strength of SOC and anisotropic Hund's coupling,
different descriptions of the multiplet structure should be employed
in the weak and strong SOC limits, \textit{viz.} \textit{LS} and \textit{jj} coupling schemes,
respectively. We investigate the implications of both the coupling
schemes on the low-energy effective $t-J$ model and calculate the
angle-resolved photoemission (ARPES) spectra using self-consistent
Born approximation. In particular, we obtain the ARPES spectra of
quasi-two-dimensional square-lattice Iridate ${\rm Sr_2IrO_4}$ in both weak
and strong SOC limits. The
differences in the limiting cases are understood in terms of the
composition and relative energy splittings of the multiplet
structure. Our results indicate that the LS coupling scheme yields
better agreement with the experiment, thus providing an indirect
evidence for the validity of LS coupling scheme for iridates. We
also discuss the implications for other metal ions with strong SOC.
\end{abstract}
\pacs{}
\maketitle

\section{Introduction}
Competition between on-site spin-orbit coupling (SOC), Coulomb repulsion
and crystal field interactions in Iridates gives rise to a plethora of unusual features.
For one of the most studied iridium-based compounds, Sr$_2$IrO$_4$, localized transport~\cite{ChengSun2016,Kim2008,Kim2009}, absence of metalization at high pressures~\cite{Haskel2012,Zocco2014} and 
emergence of an odd-parity hidden order in Rh-doped Sr$_2$IrO$_4$~\cite{ZhaoNature2016,JeongNatCom2017} were observed experimentally but are still debated from a theoretical standpoint.
On the other hand, despite many experimental indications of possible superconductivity in doped Sr$_2$IrO$_4$ -- including observation of Fermi arcs and a $d$-wave gap in electron-doped 
Sr$_2$IrO$_4$~\cite{Wang2015b, Kim2014, KimNature2016} - no direct
signatures of the superconducting state, such as zero electrical
resistance and/or Meissner effect, have been observed in these systems yet.

The ground state of Sr$_2$IrO$_4$ is believed to be an antiferromagnet
(AFM) of pseudospin $j_{\rm eff} =1/2$. The experimental low-energy magnon
dispersion is described well by the Heisenberg model with up to third
neighbor.\cite{JHKim2012}
On the theoretical side, such Heisenberg model is derived by projecting the
superexchange Kugel-Khomskii model~\cite{Oles2005} onto the spin-orbit
(SO) basis.~\cite{Jackeli2009}
However, this is a valid approach only if the virtual intermediate
doubly occupied states considered in the second order perturbation
theory
can be well approximated by the $^3T_1$, $^1T_2$, $^1E$ and $^1A_1$
basis set. Such a basis set is
an eigenbasis of the full Coulomb Hamiltonian which includes the
10\textit{Dq} crystal field as well as the Hund's coupling, but not SOC.
In other words, this approach is, strictly speaking, valid only in the limit of crystal
field and Hund's coupling much larger than SOC.
In that case, the multiplet structure of $d^4$ configuration is well described by the \textit{LS} coupling scheme.
This is indeed
the assumption made in many of the earlier works,\cite{Meetei2015, Sato2015, Kush2018, QChen2017} for instance in \onlinecite{Paerschke2017} 
while deriving the \textit{t-J}-like model of Sr$_2$IrO$_4$ to calculate the PES spectra.
The PES spectra, thus obtained, reproduces the low-energy features of
the experimental spectra remarkably well, which 
 is both interesting and intriguing. 

For materials with the
large atomic number $Z$, such as Ir, SOC is expected to be large since
it scales proportionally to $Z^4$. 
The SO
splitting in the $5d$ shell of $5d$ transition metals is $\sim 0.5$ eV.
In comparison, for transition metal (TM) atoms with partially filled $3d$ shells, such as Fe,
Ni and Co, it is one order of magnitude smaller ($\sim 0.05$ eV). For
such cases, {\it LS} coupling scheme describes the multiplet structure
well.\cite{Sobelman} For atoms with partially filled $4d$ shells, such as Ru, Rh and
Pd, the SO splitting is $\sim 0.1$ eV and there are increasing deviations
from the LS coupling scheme.\cite{Sobelman} For even heavier atoms, such
as Bi and Pb, where SO splitting is $\sim 2$ eV, the LS coupling is expected to fail.
In such cases, \textit{jj} coupling scheme would be an appropriate
choice to describe the multiplet structure.

Quantitatively, the relative strength of SOC and electron correlation is measured in terms of the ratio,~\cite{Zvezdin}
\begin{equation}
 \chi = \frac{\xi}{F_2}\,,
\end{equation}
where $\xi$ is the (single particle) on-site SOC strength and $F_2$ is a Slater integral connected
to the Slater parameter $F^{(2)}$ as $F_{2} = F^{(2)}/49$ for $d^2$
configuration.~\cite{Racah2}
Using the Racah parameters $B=420$ cm$^{-1}$ and $C=2100$ cm$^{-1}$ for Ir$^{4+}$ ion\cite{Andlauer} 
leads to $F_2 = 720\:\mathrm{cm}^{-1}$. Substituting $\xi=0.4\:\mathrm{eV}\approx3226\:\mathrm{cm}^{-1}$, we get
\begin{equation}
\chi \approx 4.5. 
\end{equation}
The \textit{LS} coupling scheme is known to be a good approximation for $\chi\lesssim 1$.~\cite{Zvezdin} Therefore, for the case of iridium the choice
of the \textit{LS} coupling scheme is questionable. 

$4d$ and $5d$ TM oxides with $J=0$ ground state has attracted a lot
of attention as it can lead to interesting effects such as excitonic
magnetism in Van-Vleck type Mott Insulators \cite{Agrestini2018arxive} or even triplon condensation and triplet
superconductivity.\cite{Horsdal2016, Chaloupka2016, Khaliullin2013, Akbari2014}
Here, caution must be exercised in the choice of the coupling scheme.
For example, the authors of Ref.~[\onlinecite{Khaliullin2013}] claim
that $4d$ and $5d$ transition metal ions with the $t_\mathrm{2g}^4$
configuration such as Re$^{3+}$, Ru$^{4+}$, Os$^{4+}$  and
Ir$^{5+}$ realize a low-spin $S = 1$ state because of relatively large
Hund's coupling % $J_\mathrm{H}$, 
and, therefore, the multiplet structure
should be calculated within the
\textit{LS} coupling scheme. While this is likely to be true for
Ru$^{4+}$ as a $4d$-element, which is, in fact, the only element discussed
in detail in 
Refs.~[\onlinecite{Chaloupka2016, Khaliullin2013, Akbari2014}], the
validity of the statement for heavier transition metal ions with
partially filled $5d$ shell is not \textit{a priori} known. 
In fact, recent analysis of 
resonant inelastic X-ray scattering data 
on double-perovskite iridium oxides with a formal valency of Ir$^{5+}$
yields SOC strength $\lambda = 0.42$ eV and Hund's coupling $J_H=0.25$
eV , suggesting {\it jj} coupling scheme to be appropriate for
Ir$^{5+}$.
\cite{Yuan2017}

One of the most prominent differences in the weak and strong SOC
strengths is the multiparticle multiplet structure which, in turn, affects the
experimentally observed features such as the PES spectra. A clear
understanding of how the low-energy description of SOC driven insulators
modifies in the weak and strong SOC limits is fundamental in developing
a satisfactory theoretical description for these systems.

In this article, therefore, we investigate the implication of the two
coupling schemes in the effective low-energy description of the ARPES spectra. 
We discuss the multiplet structures of $5d$ TM ions with the $t_{\rm 2g}^4$ 
configuration in the weak and strong SOC limits, defined by the
$LS$ and $jj$ coupling scheme, respectively. We, then, construct an
effective low-energy $t$-$J$ Hamiltonian used to describe the ARPES spectra.
For brevity, we focus on ${\rm
Sr_2IrO_4}$ to calculate the theoretical spectra within the Self Consistent
Born Approximation (SCBA) in the {\it jj} coupling scheme and make
explicit comparison with the corresponding results obtained earlier
within the {\it LS} coupling scheme~\cite{Paerschke2017} as well as the experimental
results.
This is particularly relevant in view of the fact that, despite
consensus, the validity of
the {\it LS} coupling for ${\rm Sr_2IrO_4}$ has not been established. Also, a satisfactory
theoretical description of ${\rm Sr_2IrO_4}$ is still being
developed.~\cite{JHKim2012,Cao2017}

The present work provides an indirect evidence of the validity of
the {\it LS} coupling scheme for ${\rm Sr_2IrO_4}$. More importantly, we
explicitly show the particular manifestation of the coupling schemes on
the kinetic part of a generalized \textit{t-J}-like Hamiltonian and discuss 
its ramifications. This further allows us to speculate and discuss other
scenarios where such implications could be drastic.

This article is organized as follows.
First, in Section \ref{Section:j-jL-Sintro}, we discuss the
\textit{LS} and the \textit{jj} coupling schemes within the
perturbation theory calculation of the multiplet structure. In
particular, for the case of two holes on t$_\mathrm{2g}$ shell relevant for theoretical modeling of the ARPES spectra of Iridates.  
In Section \ref{Section:j-jL-S2sites}, we discuss how the choice of the coupling scheme manifests itself in the \textit{t-J} model.
In Section \ref{Section:j-jL-Siridate}, the relevance of all these results to the calculation of ARPES spectra on Sr$_2$IrO$_4$ will be discussed.
Finally, we discuss some of the subtle issues and conclude in
Sections \ref{Sec:Discussions} \& \ref{section:jjLSconclusions},
respectively.
 
\section{Coupling schemes}
   \label{Section:j-jL-Sintro}
   
Calculating the ARPES spectral function for Sr$_2$IrO$_4$ amounts to
calculating the Green's function for the hole introduced into the AF
$j=1/2$ ground state in
the photoemission process.~\cite{Paerschke2017}

In the octahedral crystal field, the $d$ levels split into $t_\mathrm{2g}$ and $e_\mathrm{g}$ manifolds. 
There are five electrons per Ir, so effectively there is one hole
residing on the lower $t_\mathrm{2g}$ manifold. While the $t_{\rm 2g}$
manifold is composed of $d_{xy}$, $d_{xz}$, and $d_{yz}$ orbitals, the
hole carries an effective orbital momentum $l=1$ and a spin $s=1/2$ due
to orbital moment quenching.~\cite{AbragamBleaney} Due to strong on-site
SOC, the t$_\mathrm{2g}$ levels further split into $j=1/2$ doublet and
$j=3/2$ quartet and the hole occupies the lower energy
doublet.\cite{Kim2008, Jackeli2009}

\begin{figure}[!hb]
\begin{center}
\includegraphics[width=1.01\columnwidth]{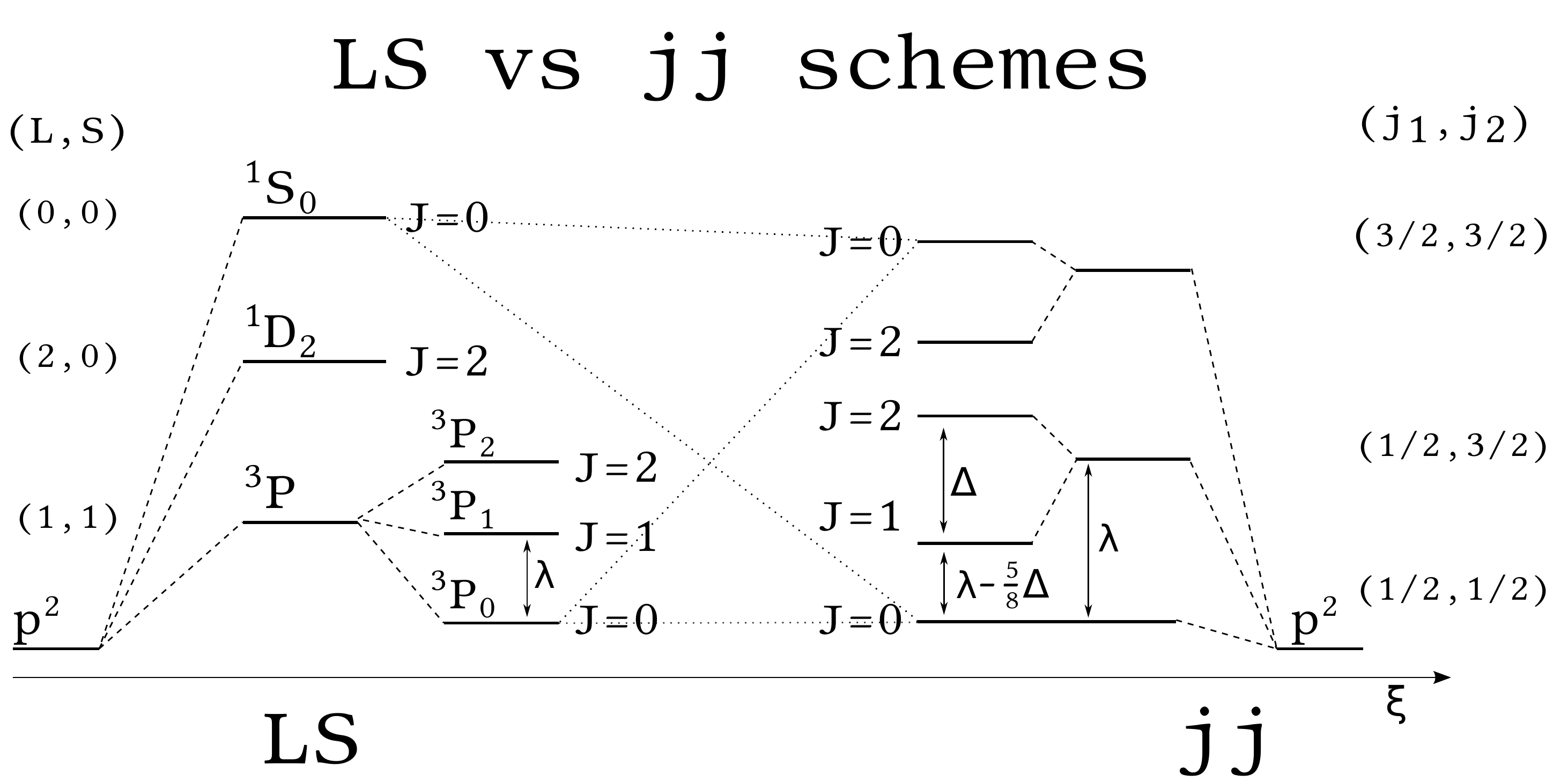}
\caption{Schematic representation of the multiplet structure for a $p^2$
    configuration in the \textit{LS} coupling scheme (left) and the
    \textit{jj} coupling scheme (right). The singlet-triplet splitting
    $\lambda = \xi/2$ where $\xi$ is the (single particle) on-site SOC strength and $\Delta$ is splitting between
    $J=1$ and $J=2$ states that depends on
    Coulomb interactions and Hunds coupling.
The mixing between $^3P_0$ and $^1S_0$ multiplets is schematically shown
    by the dotted line. 
    For comparison, the energy reference has been chosen to be 
    equal in both coupling schemes. 
\label{fig:j-jL-S}}
\end{center}
\end{figure}

Adding a hole to the Ir$^{4+}$ ion leads to the $5d^4$ configuration.
Since each hole has effective orbital momentum $l=1$ per
hole,~\cite{AbragamBleaney} the $d^4$ configuration effectively mimics
the $p^2$ configuration and we focus on the multiplet structure of the latter.
The multiplet structure depends on the coupling scheme, as shown 
in Fig.~\ref{fig:j-jL-S} and discussed in the following.

It is important to note that the need for considering either {\it
LS} or {\it jj} coupling scheme arises only for the cases when
there are more than one fermions per site. 
In such cases, the multi-particle multiplet structure differs in
the weak and strong SOC limits. For undoped Sr$_2$IrO$_4$, with only one
hole per site, both SOC and correlation effects can be treated on equal
footing.\cite{Zhong2013,Griffith}

We begin with the full Hamiltonian of a system
   \begin{equation}
      \label{Hamfull_j-jL-S}
      \mathcal{H}=\mathcal{H}_{\rm Cen}+\mathcal{H}_{\rm
      res}+\mathcal{H}_\mathrm{\rm SOC}.
    \end{equation}
Here, $\mathcal{H}_{\rm Cen}$ is the central field Hamiltonian and includes kinetic energy of all electrons,
nucleus-electron Coulomb interaction and central-symmetric part $S(r_i)$ of the Coulomb electron-electron repulsion:
   \begin{equation}
      \label{Hamcen}
      \mathcal{H}_\mathrm{Cen}=\sum_{i=1}^{N}\left(-\frac{1}{2}\nabla_{r_i}^2-\frac{Z}{r_i}+S(r_i)\right),
    \end{equation}
    where $Z$ is the atomic number of the nucleus and $N$ is the total amount of electrons in the system.     
Residual Coulomb Hamiltonian describes the angular part of the Coulomb interaction between electrons: 
   \begin{equation}
      \label{Hamres}
      \mathcal{H}_\mathrm{res}=\sum_{i>j}^{N}\frac{1}{r_{ij}}-\sum_{i=1}^{N}{S(r_i)},
    \end{equation}
    and $\mathcal{H}_\mathrm{SOC}$ describes the sum of all the on-site spin-orbit interactions
    \begin{equation}
      \label{HamSOC_j-jL-S}
      \mathcal{H}_\mathrm{SOC}=\lambda \textbf{L} \cdot
        \textbf{S}=\sum_{i=1}^{N}{\xi_i \textbf{l}_i \cdot \textbf{s}_i}.
    \end{equation}  

Eq.~(\ref{Hamfull_j-jL-S}) can be solved perturbatively, taking
$\mathcal{H}_\mathrm{Cen}$ to be the unperturbed part of the Hamiltonian.
The eigenstates of this unperturbed system are described by $\psi_\mathrm{cen}$:
\begin{equation}
 \mathcal{H}_\mathrm{Cen}\left|\psi_\mathrm{Cen}\right\rangle=E_\mathrm{Cen}\left|\psi_\mathrm{Cen}\right\rangle,
\end{equation}
and define the electronic configuration $\psi_\mathrm{Cen} = \left|n_1\:l_1, n_2\:l_2, ...\,, n_\mathrm{N} l_\mathrm{N} \right\rangle$ where $n_i$ is a principal quantum number of the $i$-th particle.

Relative strengths of $\mathcal{H}_{\rm res}$ and
$\mathcal{H}_{\rm SOC}$ dictates the order of perturbation and leads to
two different coupling schemes in the limiting cases.
If $ \mathcal{H}_\mathrm{res} > \mathcal{H}_\mathrm{SOC}$, then the strongest perturbation to the eigenstates of $\mathcal{H}_\mathrm{Cen}$ can be calculated as $\left\langle\psi_\mathrm{Cen}\left| \mathcal{H}_\mathrm{res}\right|\psi_\mathrm{Cen}\right\rangle$.
Electronic configurations then split into multiplet terms 
\begin{equation}
\psi^{LS}=\left|S\:M_S\:L\:M_L \right\rangle,
\end{equation} 
characterized by the total orbital $\textbf{L}$ and spin $\textbf{S}$
momenta. 
SOC further splits these levels and each level is now
described by the total momenta $\textbf{J}=\textbf{L}+\textbf{S}$,
as can be seen in Fig.~\ref{fig:j-jL-S}.

On the other hand, the \textit{jj} coupling scheme is applicable if
$\mathcal{H}_\mathrm{SOC} > \mathcal{H}_\mathrm{res}$, implying that
$\mathcal{H}_\mathrm{SOC}$ is the strongest perturbation 
to $\mathcal{H}_\mathrm{Cen}$. In practice, this means
that \textit{L} and \textit{S} are not good quantum numbers anymore
(i.e. they don't even form good first order approximation to the
(unknown) eigenbasis of the total Hamiltonian Eq. (\ref{Hamfull_j-jL-S}))
and the total \textbf{J} momentum has to be calculated as a sum of
individual \textbf{j} momenta characterizing each particle. 

In order to obtain the multiplet structure in the \textit{LS} (\textit{jj})
coupling scheme, an unambiguous link between the product states 
$\left|\zeta \sigma \right\rangle \left|\zeta' \sigma' \right\rangle$
(for two holes) and the final multiplet set 
$\left|S, M_S, L, M_L \right\rangle$ ($\left|j, m_{j},j',m_{j'}
\right\rangle$) should be established, where $\zeta, \zeta' = xy,yz,xz$
indicate the orbitals occupied by the holes, and $\sigma, \sigma' =
\uparrow, \downarrow$. This is followed by another basis transformation
to obtain the states in the total ${\bf J}$ momenta. In the end, the
correspondence between different J-states in the two coupling schemes
can be obtained. This involves
working with all possible configurations and could be tedious (for
details, see Appendix \ref{AppA}).

If, however, the multiplet structure in one of the coupling schemes is
known, the multiplet structure in the other scheme can be obtained
easily:
the correspondence between the multiplets $\psi^{LS}_{S\:L\:J\:M_J}$ and 
$\psi^{jj}_{j\:j'\:J\:M_J}$ obtained within the \textit{LS} and
\textit{jj} coupling schemes can, in general, be described
as~\cite{Sobelman}

\begin{equation}
\label{transition9j}
 \psi^{jj}_{j\:j'\:J\:M_J}=\sum_{L,S}{\left(ss'[S]ll'[L]J|sl[j]s'l'[j']J\right)\psi^{LS}_{S\:L\:J\:M_J}}.
\end{equation}
Since the transition between \textit{LS} and the \textit{jj} coupling
scheme is a change of the scheme of summation of four angular momenta,
the transformation coefficients in~(\ref{transition9j})
can be expressed in terms of $9j$ symbols:~\cite{Sobelman}
\begin{align}
\label{9jsymbols}
&\left(ss'[S]ll'[L]J|sl[j]s'l'[j']J\right) = \\
& \sqrt{\left(2S+1\right)\left(2L+1\right)\left(2j+1\right)\left(2j'+1\right)}\nonumber
\left\{
  \begin{array}{ccc}
  l & l' & L \\
  j & j' & J \\
  \frac{1}{2} & \frac{1}{2} & S 
  \end{array}
\right\}.
\end{align}
The values of the factor
\begin{align}
\left\{
  \begin{array}{ccc}
  l & l' & L \\
  j & j' & J \\
  \frac{1}{2} & \frac{1}{2} & S 
  \end{array}
\right\}=A\left(SLJ;\:jj'J\right)
\end{align}
are given, for example, in Table (5.23) of Ref.~[\onlinecite{Sobelman}]
or in Ref.~[\onlinecite{matsunobu1955tables}].

Let us explicitly calculate how
$\psi^{jj}_\mathrm{\frac{1}{2}\:\frac{1}{2}\:0\:0}$ transforms into the
$\psi^{LS}_{S\:L\:0\:0}$.
\begin{align}
\label{renormJLScoef}
\psi^{jj}_\mathrm{\frac{1}{2}\:\frac{1}{2}\:0\:0} &=
    \sum_{L,S}\sqrt{\left(2S+1\right)\left(2L+1\right)\left(2\cdot\frac{1}{2}+1\right)\left(2\cdot\frac{1}{2}+1\right)}\nonumber\\
&\times A\left(S\ L\ 0;\:\frac{1}{2}\ \frac{1}{2}\ 0\right)\psi^{LS}_{S\:L\:0\:0}.
\end{align}
Using table (5.23) of~[\onlinecite{Sobelman}] we calculate the values of
$A\left(S\ L\ 0;\:\frac{1}{2}\ \frac{1}{2}\ 0\right)$ and arrive at 
\begin{eqnarray}
\label{renormJLScoef2}
    \psi^{jj}_\mathrm{\frac{1}{2}\:\frac{1}{2}\:0\:0} & = & \frac{1}{\sqrt{3}}\psi^{LS}_\mathrm{0\; 0 \;0 \;0}+\sqrt{\frac{2}{3}}\psi^{LS}_\mathrm{1\; 1 \;0 \;0} \,\,\,\\
                                                      & = & \frac{1}{\sqrt{3}}\psi\left(^1S_{0, M_J=0}\right)+\sqrt{\frac{2}{3}}\psi\left(^3P_{0, M_J=0}\right) \,\,.\nonumber 
\end{eqnarray}
On the other hand, composition of the $J=1$ state remains unchanged in the
two coupling schemes. Similar to the $J=0$ states, there will also be a
mixing between higher energy states,
such as the two $J=2$ states, $^1D_2$ and $^3P_2$. However, 
the mixing between $J=2$ states is omitted from Fig.~\ref{fig:j-jL-S} for clarity.

Using Eq. (\ref{transition9j}), it is, therefore, possible to
obtain the relative composition of the multiplets in the different
coupling schemes. This has interesting consequences for the low-energy
effective $t-J$ Hamiltonian and the ARPES spectra.
More importantly, this already provides an estimate of the
relative redistribution of the spectral weight in the ARPES spectra.

\section{Manifestation of the coupling scheme in the $t-J$ model}
   \label{Section:j-jL-S2sites}

Time evolution in the Green’s function of the hole
introduced into Sr$_2$IrO$_4$ in the
photoemission process is determined by the Hamiltonian
\begin{align}
	\label{Hamd4}
{\mathcal{H}}={\mathcal{H}}_{\rm mag}+{\mathcal{H}}_{\rm SOC}+{\mathcal{H}}_{\rm{t}},
\end{align}
where ${\mathcal{H}}_{\rm mag}$ is Heisenberg Hamiltonian describing the ground state of the system which depends on first-, second- and third- neighbor exchange parameters
$J_1$, $J_2$ and $J_3$, ${\mathcal{H}}_{\rm SOC}$
describes the on-site energy of the triplet states, and
${\mathcal{H}}_{\rm{t}}$ represents the kinetic energy of the hole.\cite{Paerschke2017}

As we are interested in the low-energy description, 
in the following, we will consider only the low energy sector of the
multiplet structure consisting
of $J=0$ and $J=1$ states. % {\bf (see Section XY)}.
The $J=2$ states lie at much higher energies, approximately twice as large
as the singlet-triplet splitting \cite{Griffith, AbragamBleaney}, and are expected to
have a small contribution to the low-energy model.
We note, however, that the resulting reduced Hilbert space is not
complete. As a result, a basis transformation 
between the product state basis and the multiplet basis 
(see Appendix \ref{AppA}) in this reduced Hilbert space is not proper
and leads to issues with normalization. 
Therefore, we consider the full set of 15
configurations (microstates) formed by two holes residing on the $t_{\rm 2g}$ orbitals 
while deriving the correspondence between the multiplet structures in
the two coupling schemes. The (physical) cutoff is to be imposed only
after arriving at the final basis set which is a good approximation to
the eigenstates of the full Hamiltonian.

Detailed knowledge of the multiplet composition in
terms of the product states is also required for deriving 
the $t-J$
Hamiltonian. Therefore, in the following, we have used the explicit
transformations in the {\it jj} coupling scheme, discussed in Appendix
\ref{jjbasistr} (Eq. (\ref{jjLSHamU3}) \& (\ref{jjfinal})). Nevertheless, for completeness and for
pedagogical reasons, we provide and discuss both the schemes in detail
in Appendix \ref{AppA}.

We consider the kinetic energy part of the effective $t-J$ model ${\mathcal{H}}_{\rm{t}}$
in the two coupling schemes.
The derivation within the {\it jj} coupling scheme closely follows that
in the {\it LS} coupling scheme\cite{Paerschke2017} and consists of two main steps. We start with the application of basis
transformations Eq. (\ref{jjLSHamU3}) \& (\ref{jjfinal}) to the hopping term of \textit{t-J} model 
$\langle 5d^4_\textbf{i} \,5d^5_\textbf{j} |  \mathcal{H}_\mathrm{t}  | 5d^5_\textbf{i}\, 5d^4_\textbf{j}\rangle$ where $\mathcal{H}_\mathrm{t}$
is a general one-particle tight-binding (TB) Hamiltonian adopted from
Ref.~[\onlinecite{Paerschke2017}].
Subsequently, we apply the slave-fermion, Holstein-Primakoff, Fourier, and Bogoliubov
transformations, leading to:

\begin{align}
	    \label{Hamd4partsJJ}
	    &{\mathcal{H}}^{jj}_{\mathrm{t}}= \sum\limits_{\textbf{k}}\left(\textbf{h}_{\textbf{k}\mathrm{A}}^{\dagger}\hat{W}^{0}_{\textbf{k}}\textbf{h}^{\phantom{\dagger}}_{\textbf{k} \mathrm{A}}\!
	    +\!\textbf{h}_{\textbf{k} \mathrm{B}}^{\dagger}\hat{W}^{0}_\textbf{k} \textbf{h}^{\phantom{\dagger}}_{\textbf{k} \mathrm{B}} \right)\! +\\
	    &\! \sum\limits_{\textbf{k}, \textbf{q}} \left(  \textbf{h}_{\textbf{k-q} \mathrm{B}}^{\dagger} \hat{W}^{\mathrm{\alpha}}_{\textbf{k},\textbf{q}} \textbf{h}^{\phantom{\dagger}}_{\textbf{k} \mathrm{B}} \alpha_\textbf{q}^{\dagger}  \!+\!
  \textbf{h}_{\textbf{k-q} \mathrm{A}}^{\dagger} \hat{W}^{\mathrm{\beta}}_{\textbf{k},\textbf{q}}\textbf{h}^{\phantom{\dagger}}_{\textbf{k}\mathrm{B}} \beta_\textbf{q}^{\dagger}\!+\!\mathrm{h.c.}\right),\nonumber
  \end{align}
where $\textbf{h}^{\dagger}$ ($\textbf{h}$) represents the hole creation (annihilation)
operator written in the low-energy multiplet basis comprising of singlet
($S_{A/B}$) and
tripet states ($T_{m\,A/B}$) with $m = 0,\pm 1$ at spin-sublattices A and B:
\begin{equation}
\label{totalJcutbasisAB}
 \hat{J} = \left\{S_\mathrm{A},  T_{1 \mathrm{A}}, T_{0 \mathrm{A}}, T_{-1 \mathrm{A}}, S_\mathrm{B},  T_{1 \mathrm{B}}, T_{0 \mathrm{B}}, T_{-1 \mathrm{B}}\right\},
\end{equation}
$A$/$B$ represent the spin sublattice index accounting for the AF
order and $\alpha^\dag$($\alpha$)/$\beta^\dag$($\beta$) represents the
magnon creation (annihilation) operator on the two sublattices.

For a realistic description of the motion of charge excitation in the
AF background of $j =1/2$ pseudospins in Sr$_2$IrO$_4$, we consider tight
binding parameters obtained from density functional theory \cite{Paerschke2017} and
exchange couplings up to third
neighbor that fit the experimental magnon dispersion.
Hopping parameters are described by $8 \times 8$ matrices
due to charge excitation's internal degree of freedom and have been
denoted by $W$.
The terms 
$\hat{W}^{0}_{\textbf{k}}$ describe the nearest, next nearest, and third neighbor 
free hopping of the polaron (i.e. not coupled to magnons) and the
vertices $\hat{W}^{\mathrm{\alpha}}_{\textbf{k},\textbf{q}}$ and
$\hat{W}^{\mathrm{\beta}}_{\textbf{k},\textbf{q}}$ 
describe the polaronic hopping. They are given by

\begin{align}
     \label{W0}
 \hat{W}^{0}_\textbf{k}= \left(\begin{smallmatrix}
      \frac{3}{2}F_1 & 0 & \textrm{-}\sqrt{\frac{3}{2}}F_2 & 0 &   0 & \sqrt{\frac{3}{2}}P_2 & 0 & \textrm{-}\sqrt{\frac{3}{2}}P_1\\
      0 & F_4 & 0 & 0 &      \sqrt{\frac{3}{2}}P_1 & 0 & Q_1 & 0 \\
      \textrm{-}\sqrt{\frac{3}{2}}F_2 & 0 & F_3 & 0 &   0 & Q_2 & 0 & Q_1 \\
      0 & 0 & 0 & 0 &        \textrm{-}\sqrt{\frac{3}{2}}P_2 & 0 & Q_2 & 0 \\
      0 & \sqrt{\frac{3}{2}}P_1 & 0 & \textrm{-}\sqrt{\frac{3}{2}}P_2 &   \frac{3}{2}F_1 & 0 & \sqrt{\frac{3}{2}}F_2 & 0\\
      \sqrt{\frac{3}{2}}P_2 & 0 & Q_2 & 0  &   0 & 0 & 0 & 0  \\
      0 & Q_1 & 0 & Q_2 &    \sqrt{\frac{3}{2}}F_2 & 0 & F_3 & 0   \\
      \textrm{-}\sqrt{\frac{3}{2}}P_1 & 0 & Q_1 & 0 &   0 & 0 & 0 & F_4  \\
      \end{smallmatrix}\right), 
 \end{align}
for the free hopping matrix while the matrices containing vertices
are
\begin{align}
 \label{Walpha}
      &\hat{W}^{\alpha}_{\textbf{k},\textbf{q}}= \left(\begin{smallmatrix}
      0 & \sqrt{\frac{3}{2}}L_3 & 0 & \textrm{-}\sqrt{\frac{3}{2}}L_3 &  \frac{3}{2}Y_1 & 0 & \textrm{-}\sqrt{\frac{3}{2}}W_2 & 0\\
      \sqrt{\frac{3}{2}}L_3 & 0 & L_1 & 0 &  0 & Y_4 & 0 & W_1 \\
      0 & L_1 & 0 & L_1 &  \textrm{-}\sqrt{\frac{3}{2}}W_2 & 0 & Y_2 & 0 \\
      \textrm{-}\sqrt{\frac{3}{2}}L_3 & 0 & L_1 & 0 &  0 & W_1 & 0 & Y_3 \\
      0 & 0 & 0 & 0   & 0 & \sqrt{\frac{3}{2}}L_4 & 0 & \textrm{-}\sqrt{\frac{3}{2}}L_4 \\
      0 & 0 & 0 & 0   & \sqrt{\frac{3}{2}}L_4 & 0 & L_2 & 0  \\
      0 & 0 & 0 & 0   & 0 & L_2 & 0 & L_2 \\
      0 & 0 & 0 & 0   & \textrm{-}\sqrt{\frac{3}{2}}L_4 & 0 & L_2 & 0  \\
      \end{smallmatrix}\right), 
 \end{align}
and
 \begin{align}
 \label{Wbeta}
      &\hat{W}^{\beta}_{\textbf{k},\textbf{q}}= \left(\begin{smallmatrix}
      0 & \sqrt{\frac{3}{2}}L_4 & 0 & \textrm{-}\sqrt{\frac{3}{2}}L_4 &  0 & 0 & 0 & 0 \\
      \sqrt{\frac{3}{2}}L_4 & 0 & L_2 & 0 &  0 & 0 & 0 & 0 \\
      0 & L_2 & 0 & L_2 &  0 & 0 & 0 & 0 \\
      \textrm{-}\sqrt{\frac{3}{2}}L_4 & 0 & L_2 & 0 &  0 & 0 & 0 & 0 \\
     \frac{3}{2}Y_1 & 0 & \sqrt{\frac{3}{2}}W_2 & 0   & 0 & \sqrt{\frac{3}{2}}L_3 & 0 & \textrm{-}\sqrt{\frac{3}{2}}L_3  \\
     0 & Y_3 & 0 & W_1   & \sqrt{\frac{3}{2}}L_3 & 0 & L_1 & 0  \\
     \sqrt{\frac{3}{2}}W_2 & 0 & Y_2 & 0   & 0 & L_1 & 0 & L_1  \\
     0 & W_1 & 0 & Y_4   & \textrm{-}\sqrt{\frac{3}{2}}L_3 & 0 & L_1 & 0  \\
     \end{smallmatrix}\right), 
       \end{align} 
where ${\bf k}$-dependent hopping elements $P_i$, $Q_i$, $F_i$, and ${\bf k}$-, ${\bf q}$-dependent vertices $Y_i$, $W_i$ and $L_i$ are given in Appendix~\ref{AppD}.

Therefore, by means of Holstein-Primakoff transformation, we have
effectively mapped the complicated many-body problem onto a simpler one,
describing the motion of a polaronic quasiparticle composed of charge
excitations dressed by the $j=1/2$ magnons. This is achieved by
projecting out the interaction of magnons with each other as well as
their renormalization by the quasiparticle propagator. These
approximations comprise the well-known self-consisted Born
approximation.\cite{Martinez1991,Liu1992, Sushkov1994,Brink1998,
Shibata1999, Wang2015}

\begin{figure}[!ht]
\begin{center}
\includegraphics[width=0.95\columnwidth]{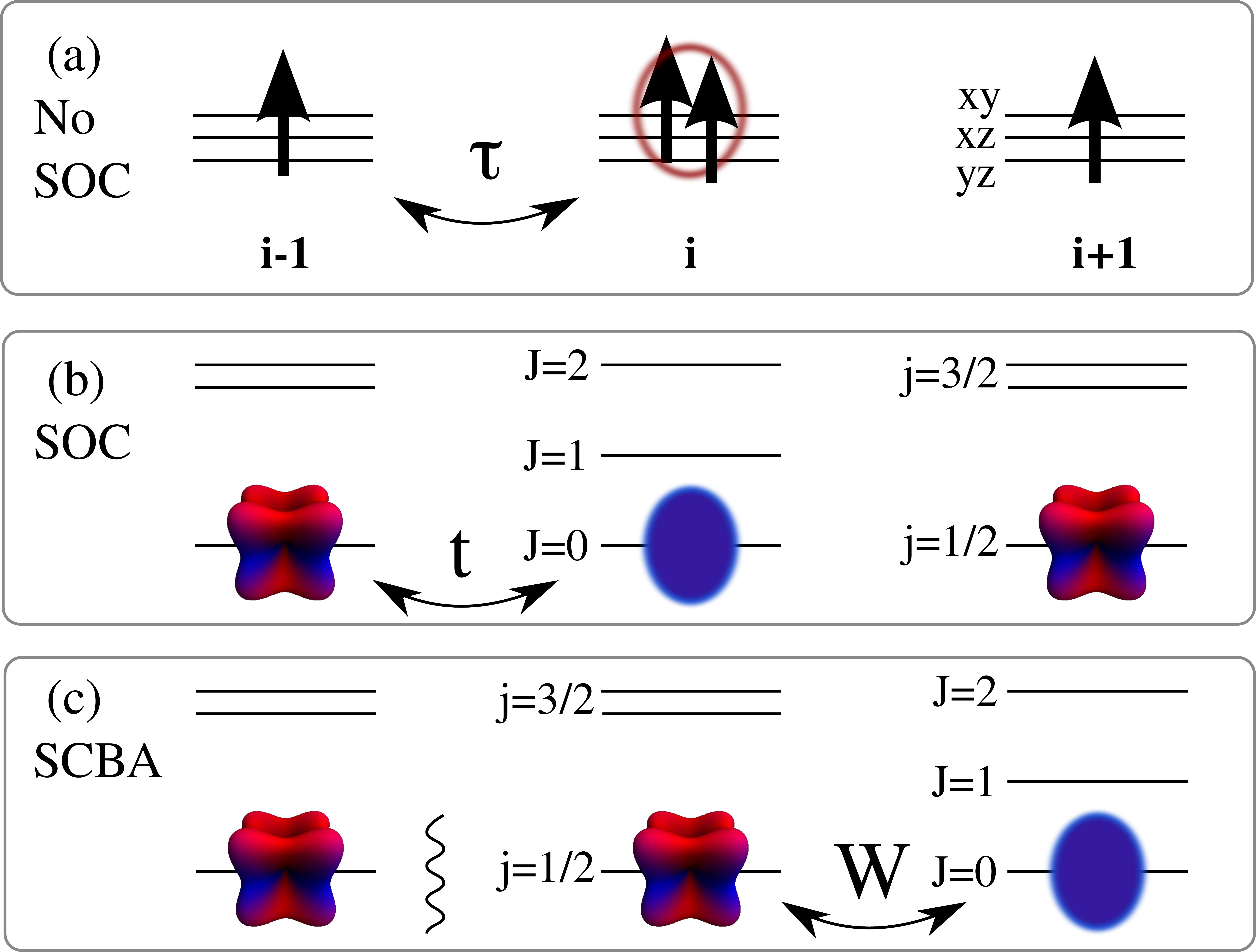}
\caption{Charge excitation on Sr$_2$IrO$_4$: (a) Without on-site
    spin-orbit coupling, there is one hole on degenerate $xy$, $yz$
    and $xz$ orbitals in the ground state on sites $i-1$ and $i+1$, and a charge excitation, i.e. a many-body state
    consisting of two holes on site $i$. (b) With on-site spin-orbit coupling, the
    ground state is described by antiferromagnetically ordered $j=1/2$ isospins and the charge excitation of a total
    momentum J possesses internal multiplet structure, calculated within
    {\it LS} or {\it jj} coupling schemes. (c) The same as (b), mapped onto
    polaronic problem. Propagation of the 
    charge excitation is described by polaron dressed by $j
    =1/2$ magnons. Upon hopping, it creates a broken antiferromagnetic
    bond of misaligned spins, shown by the wavy line.
    Here, $\tau$'s denote first-, second- and third neighbor tight binding parameters obtained from density functional theory \cite{Paerschke2017} translated
    into the many-body language in an exact-diagonalization fashion. $t$'s stand for same hopping parameters in presence of strong on-site spin-orbit coupling.
    $W$'s are derived from the $t$'s upon downfolding the model onto polaronic formalism and describe hopping parameter of the charge excitation as well as its coupling
    to magnons. $W$'s are $8 \times 8$ matrices due to charge excitation's internal degree of freedom.}
\label{cartoon2}
\end{center}
\end{figure}

A schematic description of these steps and qualitative origin of $W$ terms is shown in Fig.
\ref{cartoon2}. In the absence of SOC, 
the ground state consists of one hole per site, with a spin up or down
and occupying one of the three degenerate $t_{\rm 2g}$ orbitals, and a
charge excitation composed of two holes (site $i$, see Fig.~\ref{cartoon2}(a)).
The charge excitation is a many-body configuration $\left|a \sigma \right\rangle \left|b \sigma' \right\rangle$, 
described by total spin $S$ and orbital moment $L$. %see Fig.\ref{fig:j-jL-S}. 
Wavefunction overlap $\tau$ between neighboring
holes is material specific and can be obtained from density
functional calculations.\cite{Paerschke2017}
In the presence of SOC (Fig.~\ref{cartoon2}(b)), the ground state with
one hole per site is an antiferromagnet of $j =1/2$ pseudospins. 
The excited state, previously described by $S$ and $L$, must now be described using 
total $J$ momentum, connected to $L$ and $S$ using either \textit{LS} or \textit{jj} coupling scheme. 
Hopping parameters $t$ capture the motion of the charge
excitations and their interaction with the $j=1/2$ magnons and are derived from $\tau$'s using basis
transformations from \textit{LS} and \textit{jj} coupling schemes as
discussed in see \ref{AppA}.

Within SCBA (Fig.~\ref{cartoon2}(c)), only the non-crossing diagrams for the fermion-magnon
interaction are retained, leading to quasiparticle dressed with the
$j=1/2$ magnon (polaron). 
The motion of the polaron is now
described by the matrices $W$ which involves the coupling between the
excitation and magnons and 
are derived from $t$'s by application
of the slave-fermion, Holstein-Primakoff, Fourier, and Bogoliubov
transformations (see App. \ref{AppC1}).

The structural similarity between the resulting Hamiltonians in the two
coupling schemes (see Eq.~(\ref{Hamd4partsJJ}) above and
Eq.~(\ref{Hparts})) is evident.
However, the W-terms describing the free and polaronic hoppings are
different from the corresponding terms in the $LS$ coupling scheme.
Comparing Eqs~(\ref{W0}\,--\,\ref{Wbeta}) with Eqs.~(\ref{V0}\,--\,\ref{Vb}), one finds that changing the coupling scheme results in renormalization of free-polaron dispersion $ \hat{W}^{0}_{\textbf{k}}$
and vertices $ \hat{W}^{\mathrm{\alpha}}_{\textbf{k},\textbf{q}}$ and  $
\hat{W}^{\mathrm{\beta}}_{\textbf{k},\textbf{q}}$, in particular
for the matrix elements corresponding to the propagation of the polaron with a singlet $S_{\mathrm{A,B}}$ character. 

Thus, in the \textit{t-J} model, the coupling scheme manifests itself in the
following way:
each term of kinetic Hamiltonian~(\ref{Hamd4partsJJ}) containing
$h^\dag_{\mathrm{S\,(A,\,B)}}$
($h^{\phantom{\dag}}_{\mathrm{S\,(A,\,B)}}$) operator gets a
renormalization factor of $\sqrt{\frac{3}{2}}$ while those containing two of singlet creation (annihilation)
operators get a factor of $\frac{3}{2}$. 

The above renormalization can be explained by the mixing of the two
$J=0$ states, $^3P_0$ and $^1S_0$, as one goes from the \textit{LS} to the \textit{jj} limit. 
This mixing is shown schematically in Fig.~\ref{fig:j-jL-S} with dotted lines.
Therefore, although the choice of the coupling scheme can
not result in the change of the number of multiplets or appearance of new multiplets, it can, however, have interesting consequences for the low-energy effective model.

As evident from Eq.~(\ref{renormJLScoef2}), part of the spectral
weight of $^3P_0$ configuration in \textit{LS} coupling scheme is
transferred to higher energies in \textit{jj} coupling scheme, 
whereas some spectral weight from higher $^1S_0$ state is transferred to lower energies.
In other words, the singlet state in the \textit{jj} coupling scheme
gets some admixture of previously excited states and only
$\sqrt{\frac{2}{3}}$ of the spectral weight of the singlet derived 
in the \textit{LS} coupling scheme. This results in renormalization of
the hopping amplitudes and vertices by a factor of $\sqrt{\frac{3}{2}}$,
seen in Eqs.~(\ref{W0}\,--\,\ref{Wbeta}).
The physical consequences of this renormalization will be discussed in the next section where
the theoretical ARPES spectrum for Sr$_2$IrO$_4$ in both coupling
schemes will be compared.

\section{Influence of the coupling scheme on the spectral function of
${\rm Sr_2IrO_4}$}
   \label{Section:j-jL-Siridate}

\begin{figure}[!t]
\raggedright
\begin{minipage}{0.45\linewidth}
\subfigure{
\includegraphics[width=1.03\linewidth]{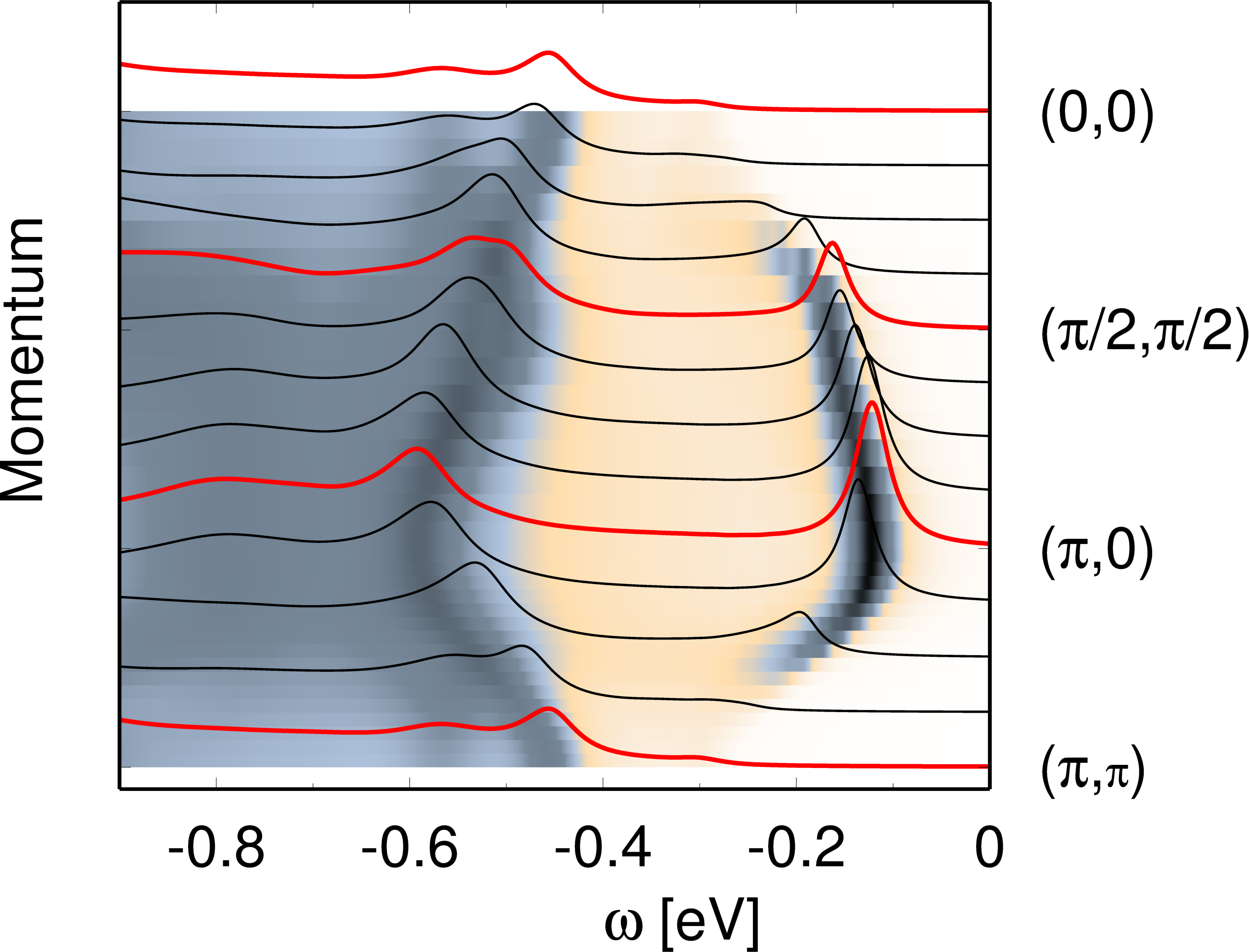}
\llap{
  \parbox[b]{2.75in}{\small{(a)}\\\rule{0ex}{1.2in}
  }}\label{fig:ARPESjj:a}%
  }
\subfigure{
\includegraphics[width=1.03\linewidth]{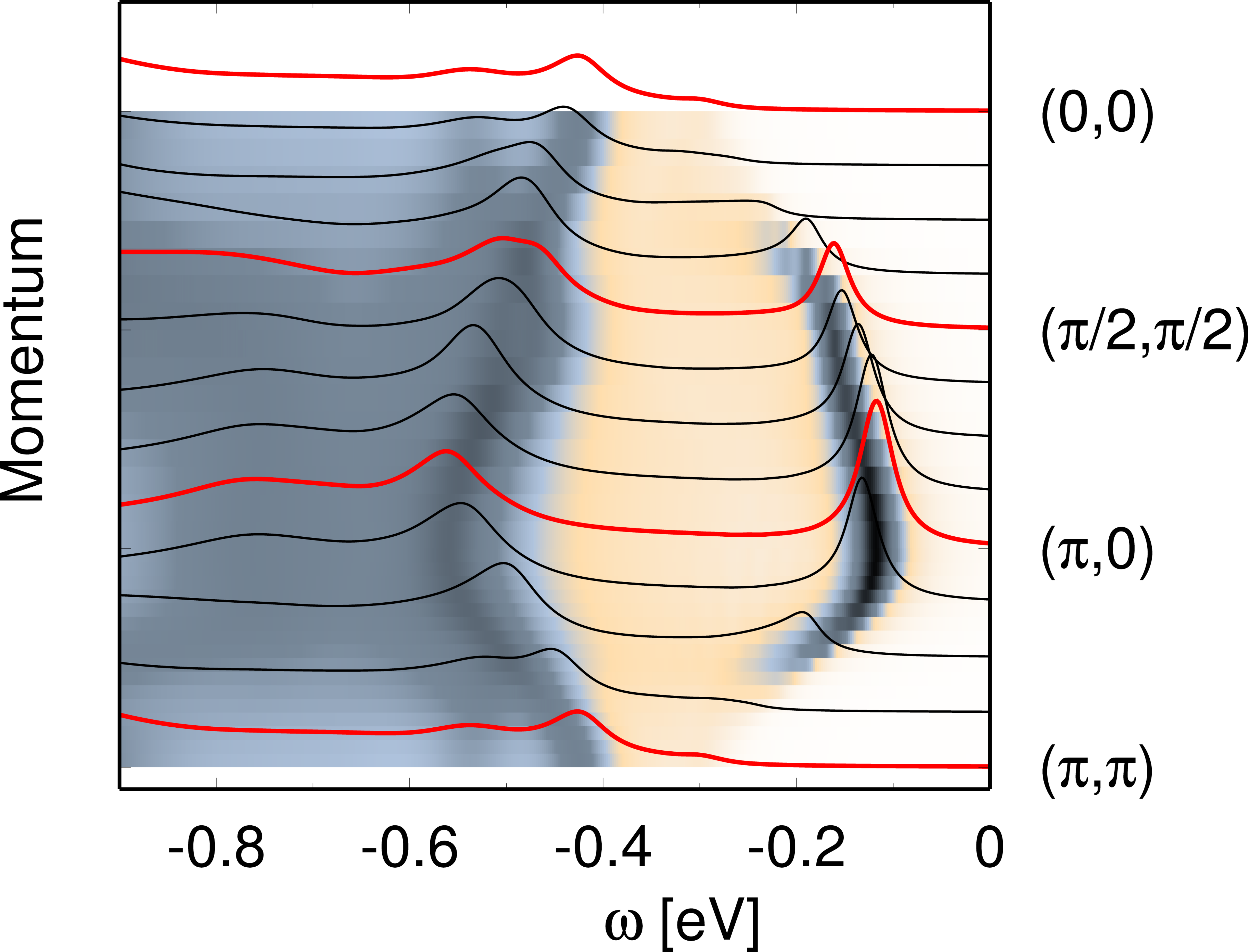}
\llap{
  \parbox[b]{2.75in}{\small{(b)}\\\rule{0ex}{1.2in}
  }}\label{fig:ARPESjj:b}%
}
\subfigure{
  \includegraphics[width=1.03\linewidth]{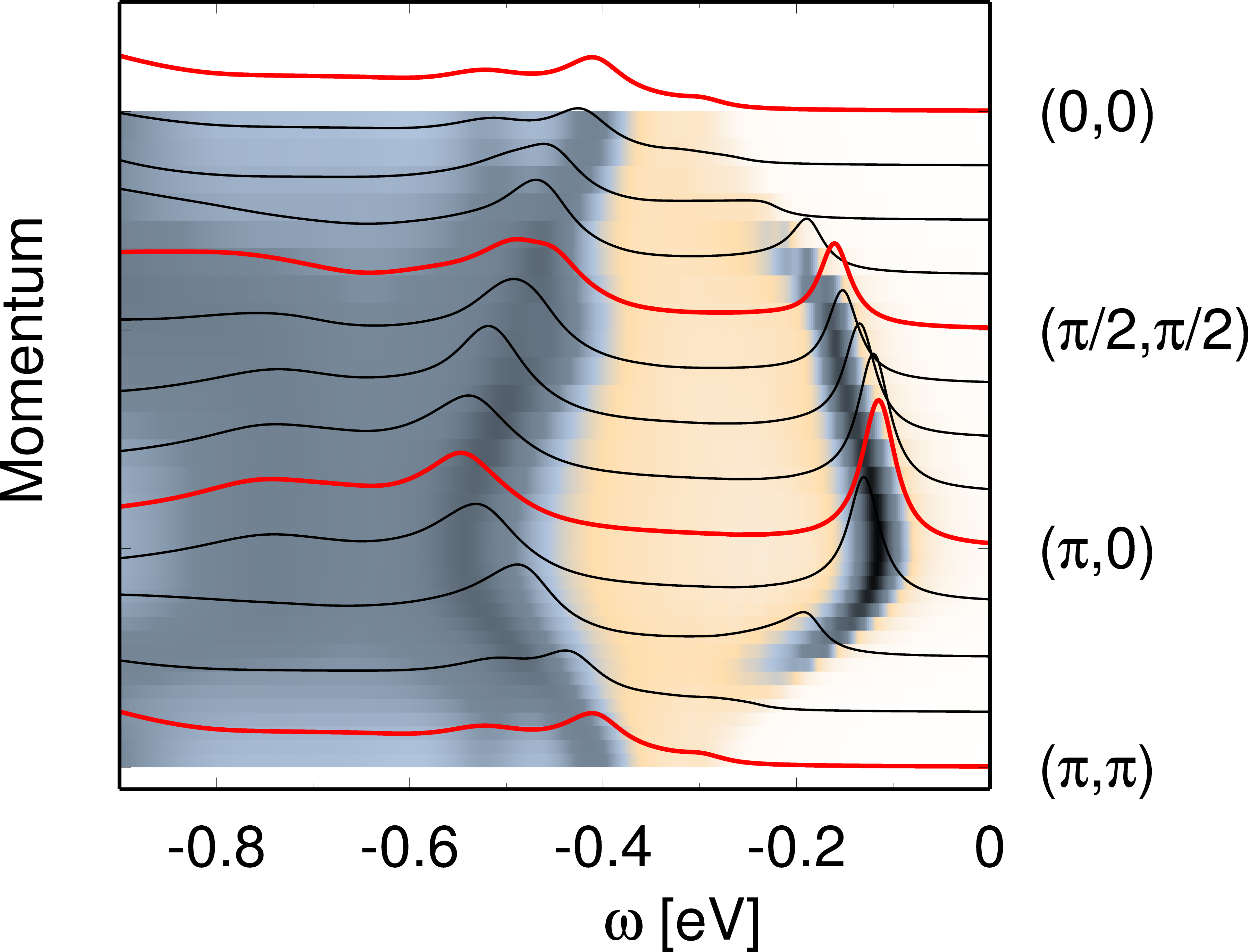}
\llap{
  \parbox[b]{2.75in}{\small{(c)}\\\rule{0ex}{1.2in}
  }}\label{fig:ARPESjj:c}%
} 
\end{minipage}
\begin{minipage}{0.45\linewidth}
\subfigure{
\includegraphics[width=0.85\linewidth]{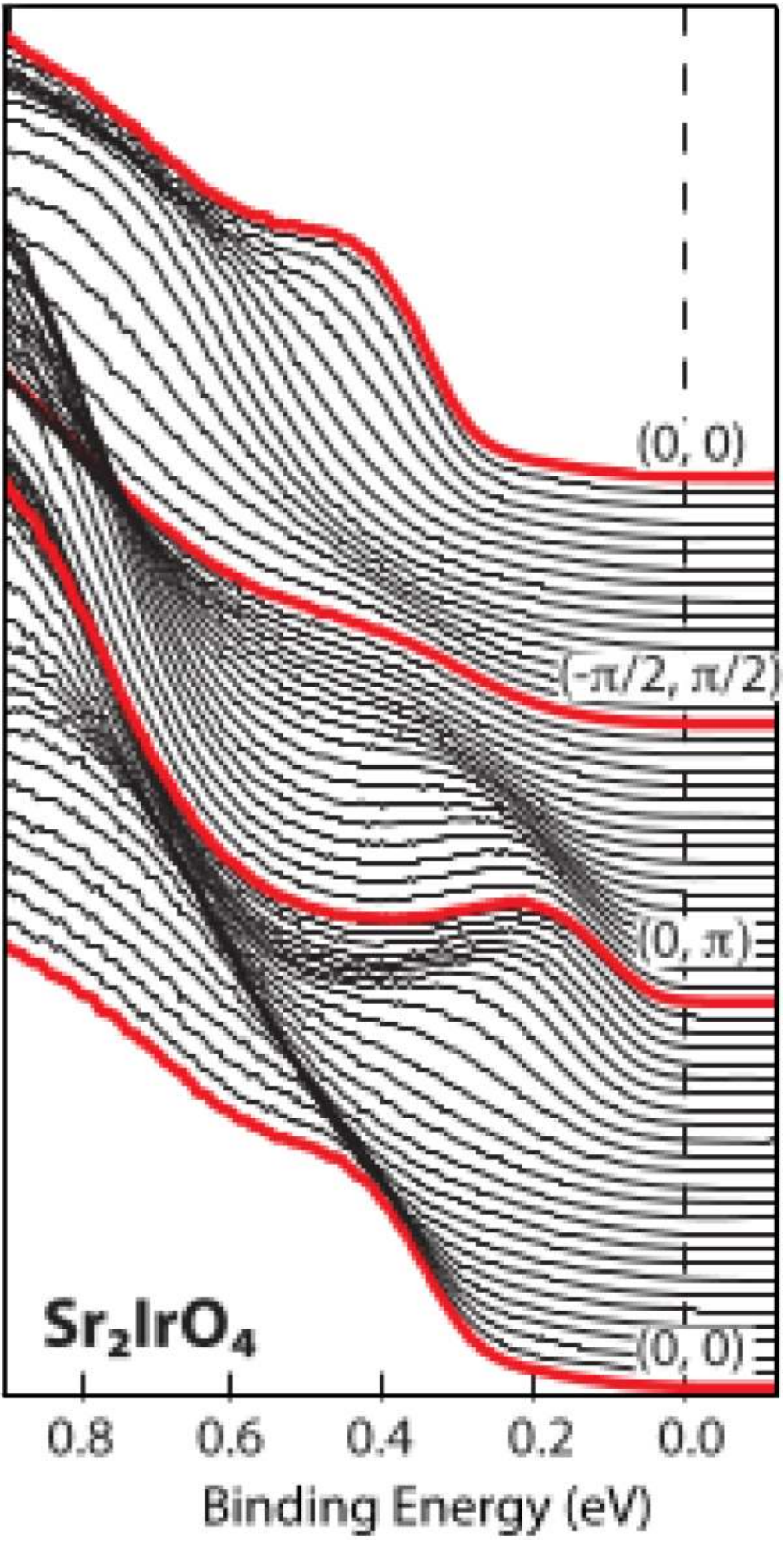}
 \llap{
  \parbox[b]{2.5in}{\small{(d)}\\\rule{0ex}{2.6in}
  }}
\label{exp}
}
\subfigure{
\includegraphics[width=1.03\linewidth]{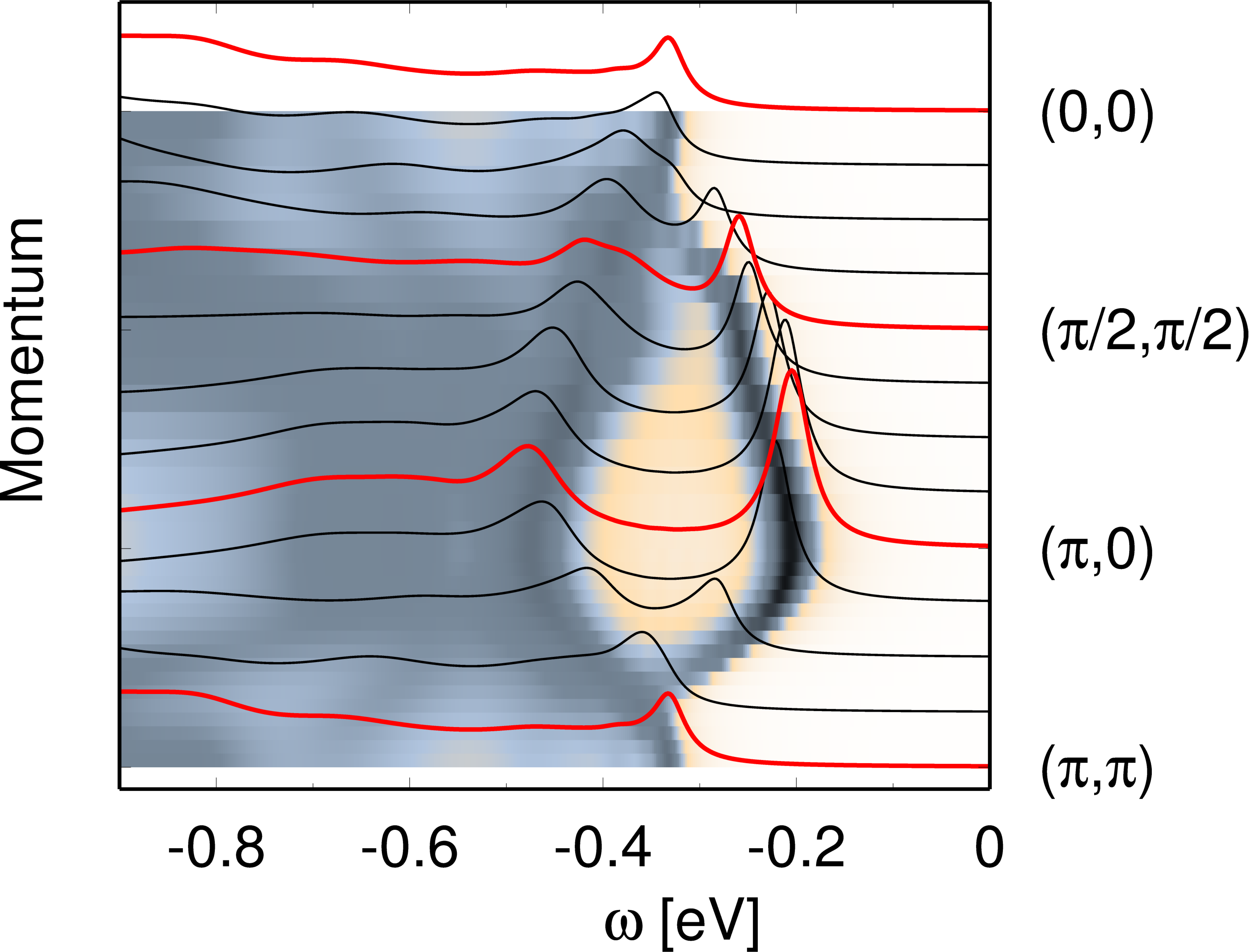}
 \llap{
  \parbox[b]{2.75in}{\small{(e)}\\\rule{0ex}{1.2in}
  }}
\label{fig:ARPESLS}
} 
\end{minipage}
\caption{PES spectral function
of the low-energy (polaronic) model developed for the quasi-two-dimensional iridates within the \textit{jj} coupling scheme and solved using the self-consistent Born approximation. 
The value of Coulomb splitting $\Delta$ varies so that singlet-triplet
    splitting:$\lambda-5/8\Delta$ is (a) $\lambda/2$, (b) $\lambda/3$, (c) $\lambda/4$. ARPES experimental data (reproduced
from Ref.[~\onlinecite{Nie2015}]) and spectral function
calculated within the \textit{LS} coupling scheme (reproduced from
Ref.~\onlinecite{Paerschke2017}) are shown for comparison in panels (d) and (e) respectively. 
Here spin-orbit coupling $\lambda=\xi/2$ where one-particle SOC
$\xi=0.382$ eV following Ref.~\onlinecite{Naturecom2014Maria}; 
hopping integrals calculated as the best fit to the density functional theory (DFT) band structure as
discussed in Ref.~\onlinecite{Paerschke2017}:
$t_1=-0.2239$ eV, $t_2=-0.373$ eV, $t'=-0.1154$ eV, $t_3=-0.0592$ eV,
$t''=-0.0595$ eV; spectra offset by (a)\,--\,(c) $E=-0.97$ eV, (e)
$E=-0.77$ eV; broadening $\delta = 0.01$ eV. 
\label{fig:ARPESjj}}
\end{figure}

Having obtained the vertices (Eqs.~(\ref{W0}\,--\,\ref{Wbeta})) describing the propagation of the polaron in Sr$_2$IrO$_4$, we calculate the Green's functions of the polaron and plot its 
spectral function whithin self consistent Born Approximation (SCBA).\cite{Paerschke2017} 
Since we don't know the exact value of splitting $\Delta$ between $\psi^{jj}_\mathrm{1\:M\:\frac{1}{2}\:\frac{3}{2}}$ and 
$\psi^{jj}_\mathrm{2\:M'\:\frac{1}{2}\:\frac{3}{2}}$ (see
Fig.~\ref{fig:j-jL-S}) which also depends on the Hund's coupling $J_\mathrm{H}$,
we consider $\Delta$ as a free parameter and perform calculations for
three values of $\Delta$ such that the singlet-triplet
splitting\footnote{The factor of 5/8 originates from the fact that the
(1/2, 3/2) state splits into a quintet ($J=2$) and a singlet ($J=1$).}
$\lambda-5/8\Delta$ takes values 
between $\lambda/2$ and $\lambda/4$ (see Fig.~\ref{fig:ARPESjj}).

There are many recent ARPES experiments revealing the shape of the iridate spectral 
functions,~\cite{Kim2008, Wang2013, delaTorre2015, Liu2015, 
Brouet2015, Cao2016, KimNature2016, Yamasaki2016,Nie2015} one of which~\cite{Nie2015} is shown on the Fig.~\ref{exp}.
 The salient features of
the spectral function are (i) lowest-energy quasiparticle peak at
($\pi$,$0$) or ($0$,$\pi$)($X$ point), 
followed by an energy gap of $\gtrsim 0.4$ eV, (ii) well defined peak at
($0$,$0$) ($\Gamma$ point), and (iii) a plateau around ($\pi/2$,$\pi/2$) ($M$ point).
While the qualitative features in all the experiments are same,
there are some quantitative differences. For instance, the splitting
between the peaks at the $X$ point and the $\Gamma$ point varies in the
range $0.15 - 0.25$ eV --- a feature crucial for explicit comparison
with the experimental data.

Comparing Fig.~\ref{fig:ARPESjj:a} and Fig.~\ref{exp}, one can see that
the low energy peaks at $M$ and $\Gamma$ points are present in the theoretical ARPES
spectra obtained within both the coupling schemes. However, as opposed
to the {\it LS} coupling scheme, for the {\it jj}
coupling scheme, the peak at the $\Gamma$ point is significantly softened in the
theoretical spectra. Furthermore, the energy gap
between the peak positions at the $\Gamma$ point and the quasiparticle
peak at $M$ is much larger for any value of singlet-triplet splitting.

\begin{figure}[!t]
 \centering
\subfigure{
\includegraphics[width=0.46\linewidth]{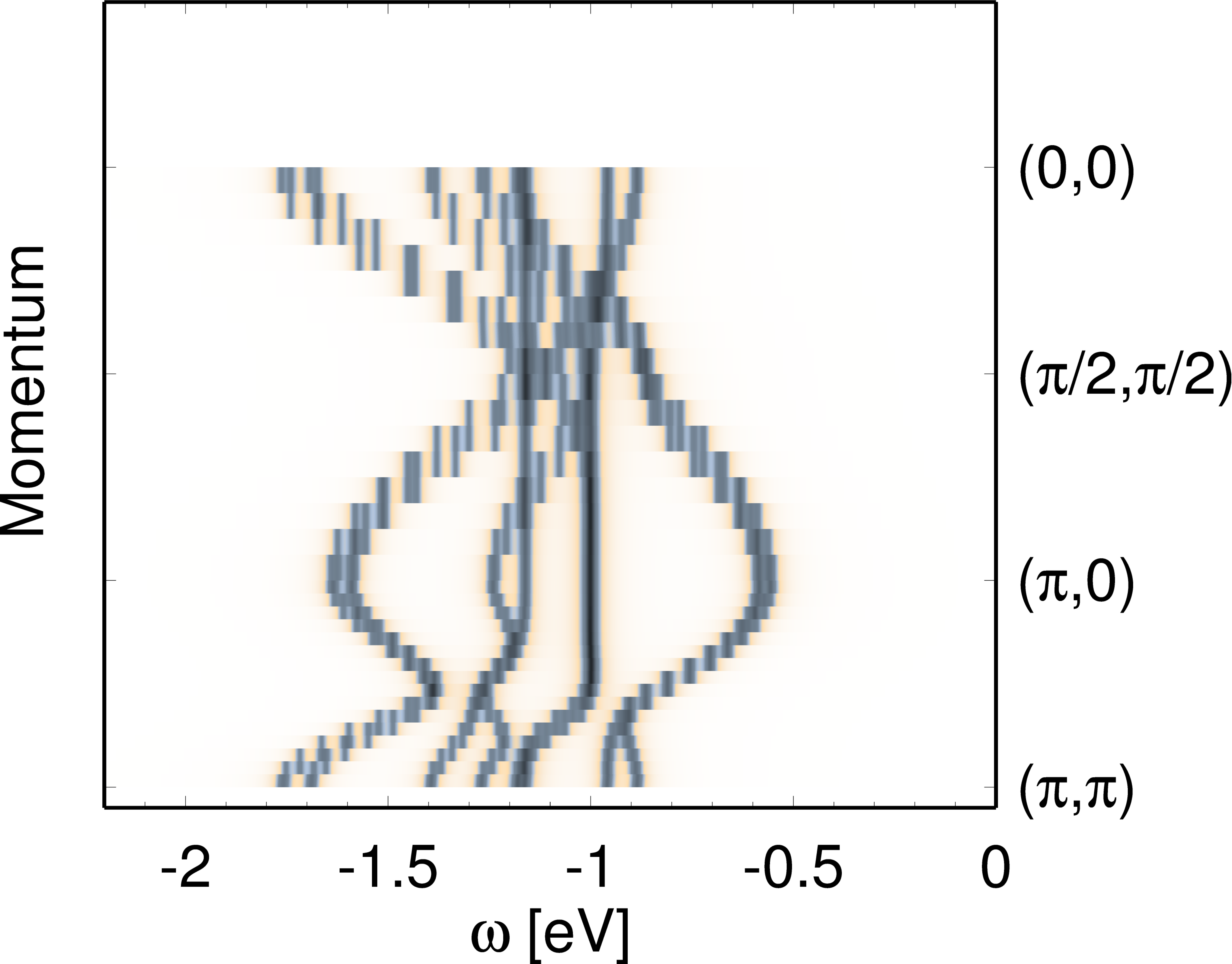}
\llap{
  \parbox[b]{2.7in}{\small{(a)}\\\rule{0ex}{1.15in}
  }}\label{fig:ARPESjj:3a}%
}
\subfigure{
\includegraphics[width=0.46\linewidth]{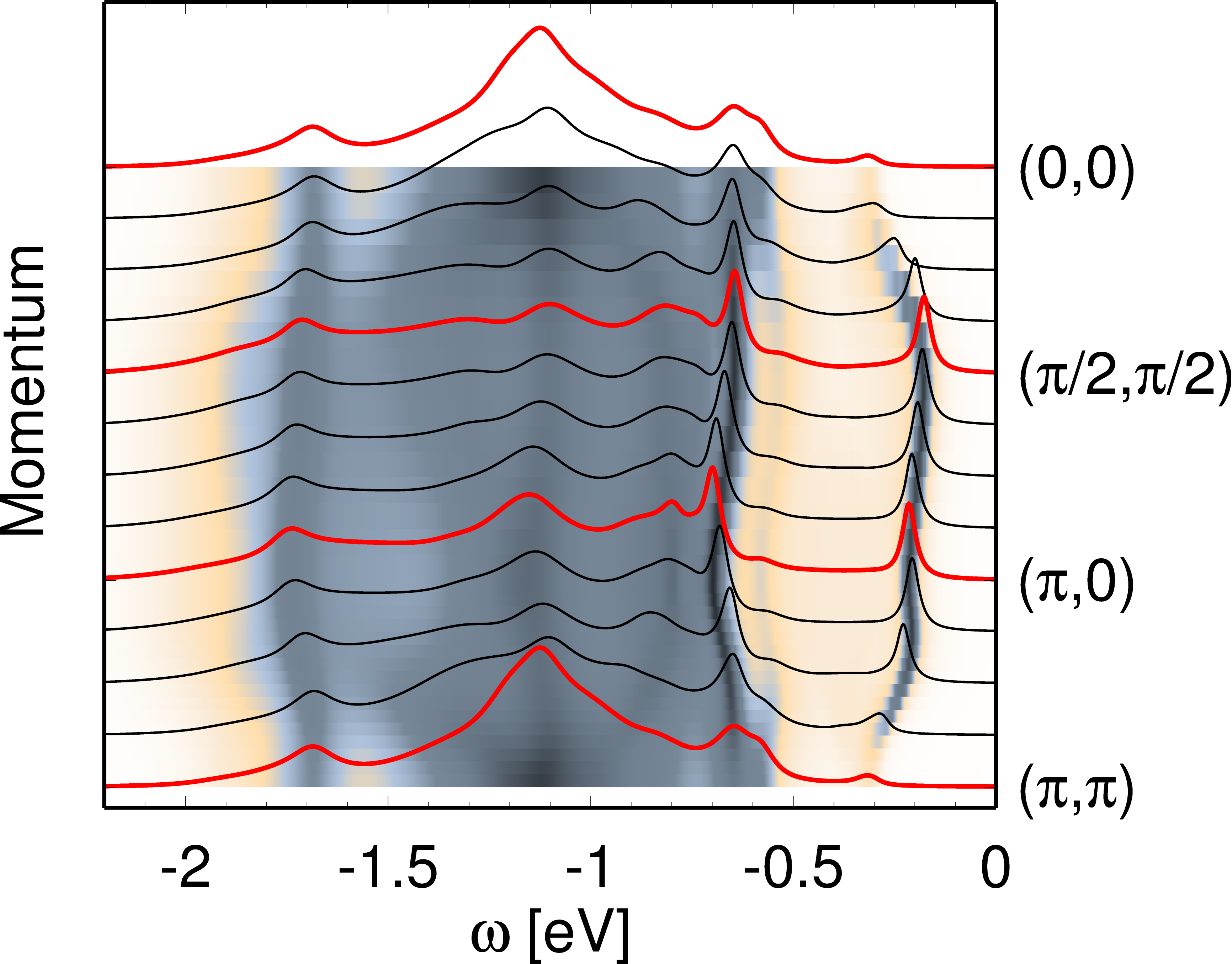}
\llap{
  \parbox[b]{2.7in}{\small{(b)}\\\rule{0ex}{1.15in}
  }}\label{fig:ARPESjj:3b}%
} 
\caption{
Free and polaronic contributions to the spectrum in Fig.~\ref{fig:ARPESjj:a}. 
      (a) Theoretical photoemission spectral function with only propagation of the hole not coupled to magnons allowed as achieved by setting $ \hat{W}^\mathrm{{\alpha}}_{\textbf{k}}=\hat{W}^\mathrm{{\beta}}_{\textbf{k}}\equiv0$.
      (b) Theoretical photoemission spectral function with only polaronic propagation via coupling to magnons allowed (i.e. no free dispersion) as achieved by 
      setting $\hat{W}^{0}_{\textbf{k}} \equiv 0$. Parameters as in
      Fig.~\ref{fig:ARPESjj}. However, note the different energy scale.
\label{fig:ARPESjj:3}} 
\end{figure}

As Coulomb $\Delta$ is varied, 
the most prominent change in the spectral function calculated within the
\textit{jj} coupling scheme is the change in the energy gap between the peak at
the $\Gamma$ point and the quasiparticle peak.  
Although the size of this gap 
depends on the value of the singlet-triplet splitting, it is not fully
determined by it. This shift of the quasiparticle peak is understood as an effect
of the renormalization of the polaronic coupling discussed earlier. 
Relatively good qualitative and quantitative agreement with the
experiment is obtained only with a
small gap
of $\lambda/4$ (Fig.~\ref{fig:ARPESjj:c}), which implies $\Delta \sim \lambda$. %one
However, as $\Delta$ becomes comparable to $\lambda$,
the \textit{LS} coupling scheme should be used, which indeed shows a
good qualitative and quantitative agreement with the experiments
(Fig.~\ref{fig:ARPESLS}).

It is interesting to note that in both {\it LS} and {\it jj} coupling
schemes, there is a reasonably sharp peak at ($\pi/2,
\pi/2$) as compared to a plateau in the experimental data.
Although the peak at ($\pi/2,\pi/2$) is suppressed in the theoretical
spectra too, owing to charge excitation scattering on magnons,
clearly, this effect is not pronounced enough. This could arise due to
overestimation of the quasiparticle spectral weight in
SCBA.~\cite{Martinez1991}
Other possibilities include 
effects beyond the approximations
made in the present study, such as hybridization of the TM
$d$ orbitals with the O $2p$ orbitals.
Such effects are known to be
important in cuprates where depending on the photon energy O $2p$ or Cu
$3d$ weights are observed in the ARPES spectra. However, for quasi-2D iridium oxides, both {\it ab-initio}
quantum chemistry calculation, as well as ARPES experiments, suggest that the charge gap is of the order of $0.5$
eV, while the Ir-O charge transfer gap is approximately 2-3 eV.~\cite{Katukuri2012, Uchida2014}
Moreover, the charge gap in the iridates is believed to be a Mott-gap~\cite{Carter2013}
that is much smaller than the charge transfer gap, putting the iridates
in the Mott-Hubbard regime. 

Yet another possibility is the role of higher lying states in the
multiplet structure. However, since a realistic description of all the
other low-energy features of the ARPES spectra is obtained for the
singlet-triplet splitting $\lambda -5\Delta/8 = 0.25 \lambda$ or in the
$LS$ coupling scheme, the relative energy difference between the $J=1$
and the $J=2$ states is $\gtrsim \lambda$. Therefore, they are expected
to have an insignificant contribution to the low-energy features.
Nevertheless, such effects can not be ruled
out completely.

\begin{figure}[!t]
 \centering
\subfigure{
\includegraphics[width=0.45\linewidth]{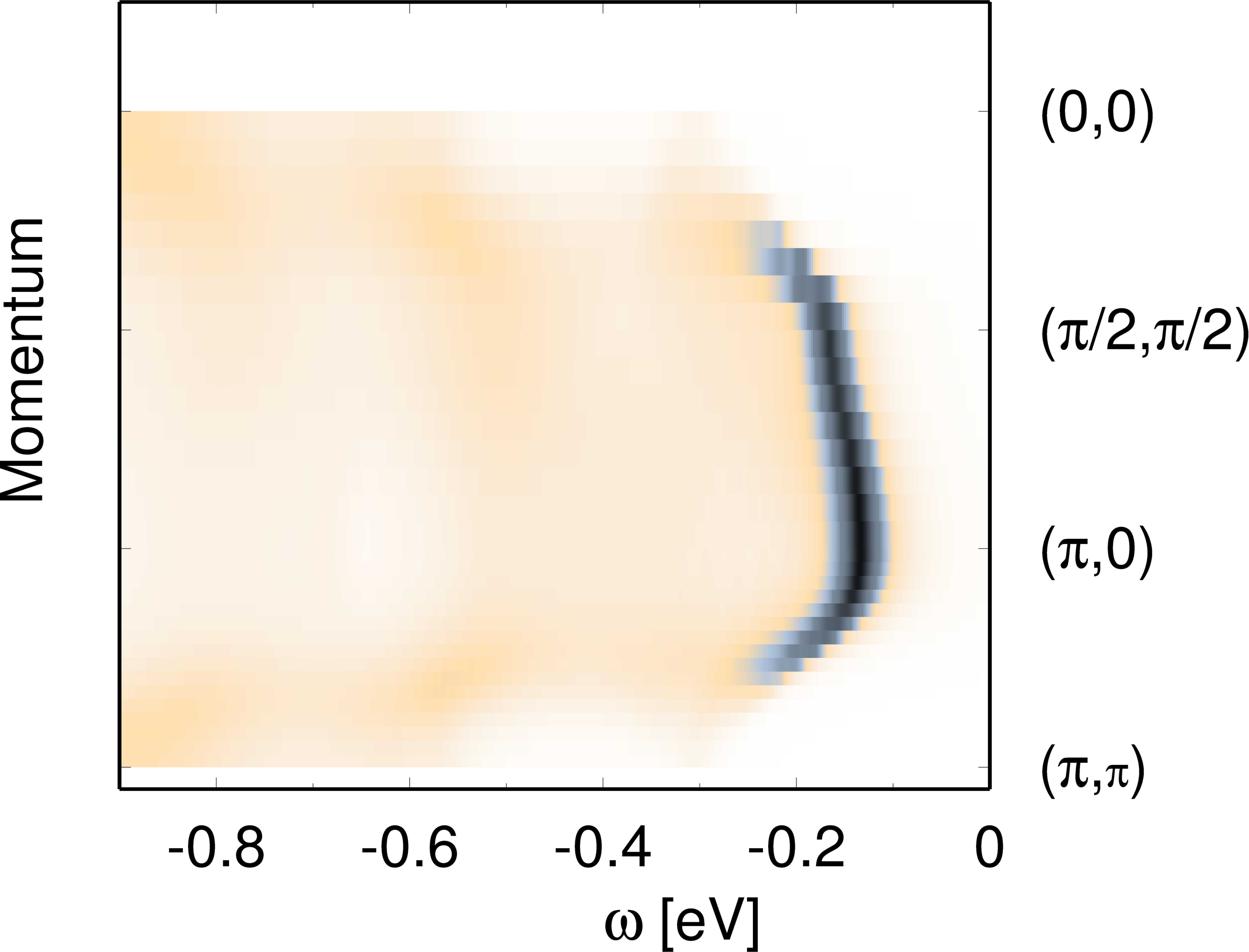}
\llap{
  \parbox[b]{1.6in}{\small{(a)}\\\rule{0ex}{1.25in}
  }}\label{fig:ARPESjj:4b}%
}
\subfigure{
\includegraphics[width=0.45\linewidth]{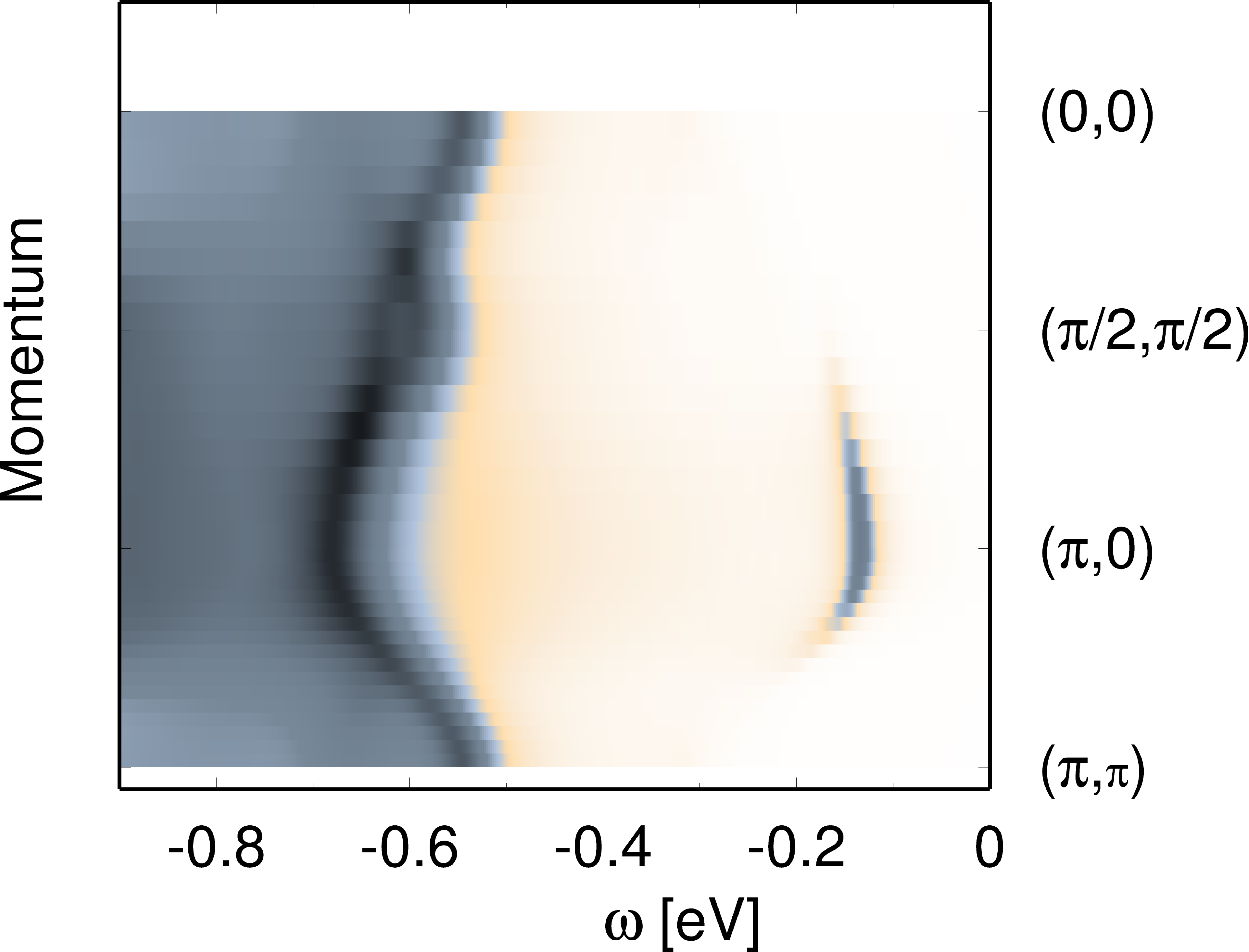}
\llap{
  \parbox[b]{1.6in}{\small{(b)}\\\rule{0ex}{1.25in}
  }}\label{fig:ARPESjj:4c}%
}
\caption{$J$-resolved theoretical photoemission spectral function of Fig.~\ref{fig:ARPESjj:a}, 
with (a) showing the $J=0$ contribution (motion of a ``singlet hole'') and (b) the $J=1$ contribution (motion of a ``triplet hole'').
\label{fig:ARPESjj:4}} 
\end{figure}

Fig.~\ref{fig:ARPESjj:3} shows the relative contributions of the
free and the polaronic part of the spectra in the {\it jj} coupling
scheme for the singlet-triplet gap equal to $\lambda$. 
Comparison with the corresponding results in the {\it LS}
coupling\cite{Paerschke2017} indicates a stronger influence
on the polaronic part of the spectra (Fig.~\ref{fig:ARPESjj:3b}) rather
than on the free part (Fig.~\ref{fig:ARPESjj:3a}). Indeed, the hole of a
singlet character has the largest contribution to the 
low-energy band (see Fig.~\ref{fig:ARPESjj:4}) and when the strength of its coupling to magnons is increased by a factor of $\frac{3}{2}$, the band gets additionally renormalized, 
thus indicating the importance of the polaronic processes. 
 
\section{Discussions}
\label{Sec:Discussions}

Most of the SO driven strongly correlated materials lie in the intermediate spin-orbit coupling regime rather than in the extreme well defined by the \textit{LS} or \textit{jj} 
coupling schemes.\cite{Sobelman}
In fact, knowledge of the composition of the low-energy states and the relative energy splittings unambiguously dictates which coupling scheme is appropriate.
In the absence of quantum
chemistry results for Ir-$d^4$ configuration,
one needs to resort to indirect verification of a suitable theoretical
model.

For ions with intermediate SOC,
ground state multiplets are in general much better captured by the \textit{LS} coupling scheme than the excited states.\cite{Zvezdin} 
For example, even for some rare-earth compounds which have $\xi\approx 1\,-\,10$, \textit{LS}
coupling usually describes the experimentally measured lowest multiplet quite well, which is however not the case for higher excited states.
For example, for Er$^{+3}$ ion, which has a value of $\xi\approx 5.53$
close to Ir, the ground-state wave function is given by~\cite{Zvezdin}
\begin{equation}
\label{eqZvezdin1}
 |\psi_{GS}\rangle=0,982|^4I\rangle-0,186|^2K\rangle\approx|^4I_{15/2}\rangle\,.
\end{equation}
i.e. the ground state is indeed well described by the \textit{LS} coupling scheme. However, already for the highest exited multiplet in the same term we have
\begin{align}
\label{eqZvezdin2}
 &|\psi_{1}\rangle=0,627|^4I\rangle-0,416|^2K\rangle-0,342|^2G\rangle -\\
 &-0,219|^2H\rangle+0,276|^2G'\rangle+0.438|^2H'\rangle\,.\nonumber
\end{align}
We see that the multiplet $^4I$, which according to the \textit{LS} coupling scheme should describe $ |\psi_{1}\rangle$, has in fact only $39$\%  
contribution in the corresponding excited wave function.~\cite{Zvezdin}

It is also important to note that, in the case of Ir, the first excited state $^3P_1$ is not affected by the coupling scheme choice 
as there exist a unique $J=1$ state.  
However, this is not the case for, i.e., $p^3$ and $p^4$ configurations. In $p^3$ configuration, two lowest multiplets, $^4S_{\frac{3}{2}}$ and $^4D_{\frac{3}{2}}$, can in general
mix with each other as well as with higher lying $3P_{\frac{3}{2}}$. In the $p^4$ configuration, where the order of some states is inverted as compared to the $p^2$ configuration, 
the first two excited multiplets $^3P_0$ and $^3P_2$ do change places
upon going from one coupling scheme to another,\cite{Sobelman} probably
rendering more pronounced effects in the theoretical description.
One can, in general, expect much bigger ramifications of the coupling scheme choice in the cases where the composition of the excited states are different as well since under the same values of SOC they
usually do get renormalized much more than the ground state, as exemplified by Eqs.~(\ref{eqZvezdin1})\,--\,(\ref{eqZvezdin2}).

Naturally, the same renormalization effect discussed in the present
work would also be observed for an electron in the material with $t^1_\mathrm{2g}$ configuration in the ground state and strong on-site SOC for any geometry 
and choice of hopping parameters. For example, 
deriving a \textit{t-J} model for a honeycomb iridates with one hole which forms the many-body $d^4$ configurations as well, one would get the same renormalization of the kinetic Hamiltonian
when going from \textit{LS} to \textit{jj} limit,
even though the motion of free charge on the honeycomb lattice is described by a completely different TB model:  the hoppings between different orbitals are much larger than 
the hoppings between the same ones
and they are moreover strongly bond-dependent.\cite{Foyevtsova2013}

For the present case, employing the DFT-based TB parameters accounts
for the crystal field effects and distortions such as octahedra rotation. We note, However, considerable
differences from the present case are expected in strong distortions,
{\it e.g.} under pressure, due to additional
mixing of the states,\cite{Bogdanov2015} and, even more importantly, the renormalization of the Clebsch-Gordan coefficients.\cite{Jackeli2009} 

Furthermore, the fact that multiplet structure of Ir$^{5+}$ can be so
well described by \textit{LS} coupling scheme also suggests that the
superexchange model for Sr$_2$IrO$_4$ can be derived by simply projecting the
Kugel-Khomskii model\cite{Oles2005} onto the spin-orbit coupled basis as
done in e.g. Ref. [\onlinecite{Jackeli2009}].

\section{Conclusions}
\label{section:jjLSconclusions}

In conclusion, we have studied the ARPES spectra for quasi-2D square
lattice iridates in weak and strong SOC strengths where the multiplet
structures are well defined by different coupling schemes. 
Specifically, we have studied how the choice of the coupling scheme can
influence the multiplet structure and consequently the low-energy
effective model for ${\rm Sr_2IrO_4}$, effectively described by $p^2$ configuration.
We have shown that for a \textit{t-J}-like model for Sr$_2$IrO$_4$, the \textit{jj} coupling scheme induces
renormalization of the vertices in the kinetic part of the Hamiltonian and prominent changes in the spectral function calculated within SCBA.
We have compared the spectra calculated in both coupling schemes to
the experimental ARPES data. 
Interestingly, despite large SOC, we find much better agreement to the experiment for the model derived
within the \textit{LS} coupling scheme. We argue that just as well as for many rare-earth compounds, which have comparable SOC strength, the spin-orbit coupling, albeit strong, is yet weak enough
to allow for a successful description of the ground state in the framework of the \textit{LS} coupling scheme. 

For other electronic configurations, such as  $p^3$ or $p^4$, where all of the low-energy multiplets are renormalized as we go from \textit{LS} to \textit{jj} coupling scheme \cite{Rubio1986}, more dramatic
consequences are expected in the theoretical ARPES spectra.

Although, the choice of the coupling scheme and the effective low-energy model can be guided by the knowledge of the composition and relative energy splittings of the multiplets, in the absence of such 
experimental and/or quantum chemistry studies, the validity of the same must be ascertained.

\section{Acknowledgements}
Authors thank Manuel Richter, Klaus Koepernik, Krzysztof Wohlfeld,
Jeroen van den Brink, Flavio Nogueira, Dmytro Inosov and
Robert Eder for helpful suggestions and discussions.
RR acknowledges financial support from the European Union (ERDF) and the Free State of Saxony via the ESF project 100231947 (Young Investigators Group Computer Simulations
for Materials Design - CoSiMa.) 

\appendix

\section{Multiplet Structure}
\label{AppA}
\subsection{\textit{LS} coupling scheme}
\label{perturbationLS}

To calculate the multiplet structure of $p^2$ configuration in \textit{LS} coupling scheme as used in Ref.~[\onlinecite{Paerschke2017}], one has 
 to establish an unambiguous link between the single particle states $\left|\zeta \sigma
 \right\rangle,\left|\zeta' \sigma' \right\rangle$ (for two holes) and the final multiplet set
$\left|S, M_S, L, M_L \right\rangle$ where $\zeta,\zeta'= xy,yz,xz$
indicate the orbitals occupied by the holes, and $\sigma, \sigma' =
\uparrow, \downarrow$. 
This is done in the following way.

First, one has to make a basis transformation from the real space basis
$\left|\zeta \sigma\right\rangle$ to the single-particle
%$\left|\alpha \sigma\right\rangle$ to the single-particle
states in the $Y_{lm}$ basis $\left|l s m_l m_s\right\rangle$.\cite{Jackeli2009}
Secondly, for multi-particle configurations, one must construct the
basis transformation from the product states to states described by total {\it L} and {\it S}.
In principle, one can use Clebsch-Gordan coefficients
(CGCs).
However, there is a caveat:
Clebsch-Gordan tables are formulated for summation of momenta of two
\textit{inequivalent} electrons. So, if we want to sum spins
$\textbf{s}_1$ and $\textbf{s}_2$ of two electrons, they must be
distinguishable. If they were on two different sites, then the position would suffice. However, if they are on the same site, as in our case, 
the multi-particle state can be obtained correctly by CGCs only if they
reside on different orbitals.
Bearing this in mind we avoid using CGCs for two-particle configurations
and instead perform moment summation using the \textit{high weight
decomposition method}, discussed in detail in Appendix \ref{AppB}. 

As a result, we can construct the matrix $U_1$ that transforms the Hamiltonian $\mathcal{H}_{ls}$ from the product 
state basis $\left|l\:s\:m_l\:m_s\right\rangle\left|l'\:s'\:m'_l\:m'_s\right\rangle$ 
to the total spin and orbital momentum basis $\left|L\:S\:M_L\:M_S\right\rangle$:
\begin{equation}
 \mathcal{H}^{LS}_{LS\mathrm{-basis}}=U_1^\dag \mathcal{H}_{ls\mathrm{-basis}} U_1^{\phantom{\dagger}},
\end{equation}
where
\begin{align}
\label{U1matrix}
       U_1 =\left(\begin{smallmatrix}
 0 & 1 & 0 & 0 & 0 & 0 & 0 & 0 & 0 & 0 & 0 & 0 & 0 & 0 & 0 \\
 0 & 0 & 0 & 1 & 0 & 0 & 0 & 0 & 0 & 0 & 0 & 0 & 0 & 0 & 0 \\
 0 & 0 & 0 & 0 & 0 & 0 & 0 & 0 & 0 & 0 & 1 & 0 & 0 & 0 & 0 \\
 0 & 0 & \frac{1}{\sqrt{2}} & 0 & 0 & \frac{1}{\sqrt{2}} & 0 & 0 & 0 & 0 & 0 & 0 & 0 & 0 & 0 \\
 0 & 0 & 0 & 0 & \frac{1}{\sqrt{2}} & 0 & 0 & \frac{1}{\sqrt{2}} & 0 & 0 & 0 & 0 & 0 & 0 & 0 \\
 0 & 0 & 0 & 0 & 0 & 0 & 0 & 0 & 0 & 0 & 0 & \frac{1}{\sqrt{2}} & \frac{1}{\sqrt{2}} & 0 & 0 \\
 0 & 0 & 0 & 0 & 0 & 0 & 1 & 0 & 0 & 0 & 0 & 0 & 0 & 0 & 0 \\
 0 & 0 & 0 & 0 & 0 & 0 & 0 & 0 & 1 & 0 & 0 & 0 & 0 & 0 & 0 \\
 0 & 0 & 0 & 0 & 0 & 0 & 0 & 0 & 0 & 0 & 0 & 0 & 0 & 1 & 0 \\
 1 & 0 & 0 & 0 & 0 & 0 & 0 & 0 & 0 & 0 & 0 & 0 & 0 & 0 & 0 \\
 0 & 0 & \frac{1}{\sqrt{2}} & 0 & 0 & -\frac{1}{\sqrt{2}} & 0 & 0 & 0 & 0 & 0 & 0 & 0 & 0 & 0 \\
 0 & 0 & 0 & 0 & \frac{1}{\sqrt{6}} & 0 & 0 & -\frac{1}{\sqrt{6}} & 0 & \sqrt{\frac{2}{3}} & 0 & 0 & 0 & 0 & 0 \\
 0 & 0 & 0 & 0 & 0 & 0 & 0 & 0 & 0 & 0 & 0 & \frac{1}{\sqrt{2}} & -\frac{1}{\sqrt{2}} & 0 & 0 \\
 0 & 0 & 0 & 0 & 0 & 0 & 0 & 0 & 0 & 0 & 0 & 0 & 0 & 0 & 1 \\
 0 & 0 & 0 & 0 & \frac{1}{\sqrt{3}} & 0 & 0 & -\frac{1}{\sqrt{3}} & 0 & -\frac{1}{\sqrt{3}} & 0 & 0 & 0 & 0 & 0 \\
\end{smallmatrix}
\right),\nonumber
\end{align}
and the product state basis $\left|l\:s\:m_l\:m_s\right\rangle\left|l'\:s'\:m'_l\:m'_s\right\rangle$ is defined as 
 \begin{eqnarray}
\label{lsbasis}
 \hat{a}&=&\{
 |1\:1\:0\:0\:0\:0\rangle,
 |1\:0\:1\:0\:0\:0\rangle,
 |1\:0\:0\:1\:0\:0\rangle,
 |1\:0\:0\:0\:1\:0\rangle,\nonumber\\
    &  &|1\:0\:0\:0\:0\:1\rangle,
 |0\:1\:1\:0\:0\:0\rangle,
 |0\:1\:0\:1\:0\:0\rangle,
 |0\:1\:0\:0\:1\:0\rangle, \nonumber \\
    &  &|0\:1\:0\:0\:0\:1\rangle,
 |0\:0\:1\:1\:0\:0\rangle,
 |0\:0\:1\:0\:1\:0\rangle,
 |0\:0\:1\:0\:0\:1\rangle, \nonumber \\
    &  &|0\:0\:0\:1\:1\:0\rangle,
 |0\:0\:0\:1\:0\:1\rangle, 
 |0\:0\:0\:0\:1\:1\rangle \}^\intercal,
 \end{eqnarray}
where 1(0) represents the (un-)occupied single particle state
of the Hilbert space spanned by $\left|m_l m_s\right\rangle = 
\left\{\left|1 \uparrow \right\rangle, \left|1 \downarrow \right\rangle, \left|0 \uparrow \right\rangle, \left|0 \downarrow \right\rangle, \left|-1 \uparrow \right\rangle, 
\left|-1 \downarrow \right\rangle \right\}^\intercal$.
The multiplet basis $\left|S\:M_S\:L\:M_L \right\rangle$ is defined as
\begin{eqnarray}
\label{multipletbasis}
    \hat{A}&=&\{\left|1\: 1\: 1\: 1\right\rangle, \left|1\:1\:1\:0\right\rangle, \left|1\: 1\: 1\: {-1}\right\rangle, \left|1\: 0\: 1\: 1\right\rangle, \left|1\: 0\: 1\: 0\right\rangle, \nonumber\\
    & &\left|1\: 0\: 1\: {-1}\right\rangle, \left|1\: {-1}\: 1\: 1\right\rangle, \left|1\: {-1}\: 1\: 0 \right\rangle,
\left|1\:  {-1}\: 1\: {-1}\right\rangle, \left|0\: 0\: 2\: 2\right\rangle,\nonumber\\ 
    & &\left|0\: 0\: 2\: 1\right\rangle, \left|0\: 0\: 2\: 0\right\rangle, \left|0\: 0\: 2\: {-1}\right\rangle, 
\left|0\: 0\: 2\: {-2}\right\rangle, \left|\: 0\: 0\: 0\:
    0\right\rangle\}^\intercal,\nonumber \\
\end{eqnarray}
so that 
\begin{equation}
 \hat{A}=U_1 \hat{a}.
\end{equation}

Upon employing this transformation, we have effectively taken Hamiltonian~(\ref{Hamres}) that defines the first-order corrections to the eigenstates of the system into account. 

According to
the Hund's rules, the state with the lowest energy is the one with the highest multiplicity and the highest possible \textbf{L},
{\it i.e.} in the first approximation the ground state is the nine-fold degenerate $^3P$ multiplet. 

To account for further perturbation on the system induced by the strong
on-site spin-orbit coupling (Eq. (\ref{HamSOC_j-jL-S})), we perform a
basis transformation to obtain the total
\textbf{J} momenta.
To build a low-energy effective model, we truncate the Hilbert space down to the high spin $^3P$ states only. 
Since total spin $S=1$ and orbital momenta $L=1$ are distinguishable by
their nature, we can simply use the CG coefficients to sum them up,
leading to
\begin{equation}
\label{lsfinal}
 \mathcal{H}^{LS}_{J\mathrm{-basis}}=U_2^\dag \mathcal{H}^{LS}_{LS\mathrm{-basis}} U_2^{\phantom{\dagger}}\,,
\end{equation}
where, $U_2$ is
\begin{align}
\label{U2matrix}
       U_2 =
 \left(
\begin{array}{ccccccccc}
 0 & 0 & \frac{1}{\sqrt{3}} & 0 & -\frac{1}{\sqrt{3}} & 0 & \frac{1}{\sqrt{3}} & 0 & 0 \\
 0 & \frac{1}{\sqrt{2}} & 0 & -\frac{1}{\sqrt{2}} & 0 & 0 & 0 & 0 & 0 \\
 0 & 0 & \frac{1}{\sqrt{2}} & 0 & 0 & 0 & -\frac{1}{\sqrt{2}} & 0 & 0 \\
 0 & 0 & 0 & 0 & 0 & \frac{1}{\sqrt{2}} & 0 & -\frac{1}{\sqrt{2}} & 0 \\
 1 & 0 & 0 & 0 & 0 & 0 & 0 & 0 & 0 \\
 0 & \frac{1}{\sqrt{2}} & 0 & \frac{1}{\sqrt{2}} & 0 & 0 & 0 & 0 & 0 \\
 0 & 0 & \frac{1}{\sqrt{6}} & 0 & \sqrt{\frac{2}{3}} & 0 & \frac{1}{\sqrt{6}} & 0 & 0 \\
 0 & 0 & 0 & 0 & 0 & \frac{1}{\sqrt{2}} & 0 & \frac{1}{\sqrt{2}} & 0 \\
 0 & 0 & 0 & 0 & 0 & 0 & 0 & 0 & 1 \\
\end{array}
\right) \,\,.
\end{align}
$\mathcal{H}^{LS}_{J\mathrm{-basis}}$ is written in the spin-orbit coupled basis
\begin{equation}
\label{totalJfullbasis}
 \hat{J} = \left\{S, T_1, T_0, T_{-1}, M_2, M_1, M_0, M_{-1}, M_{-2} \right\}^\intercal,
\end{equation}
which consists of the lowest $J=0$ singlet $S$, the higher $J=1$ triplets $T_m$ ($m=-1,0,1$, split by energy $\lambda$ from the singlet state) and $J=2$ quintets.~\cite{Paerschke2017}

To arrive at the final effective low-energy model we further truncate the Hilbert space and reduce the basis set to the two lowest multiplets $^3P_0$ and $^3P_1$ (see Fig \ref{fig:j-jL-S}):
\begin{equation}
\label{totalJcutbasis}
 \hat{J} = \left\{S,  T_1, T_0, T_{-1} \right\}^\intercal.
\end{equation} 

\subsection{\textit{jj} coupling scheme}
\label{jjbasistr}

The \textit{jj} coupling scheme is applicable if
$\mathcal{H}_\mathrm{SOC} > \mathcal{H}_\mathrm{res}$, implying that
$\mathcal{H}_\mathrm{SOC}$ is the strongest perturbation 
to $\mathcal{H}_\mathrm{Cen}$. In practice, this means
that \textit{L} and \textit{S} are not good quantum numbers anymore
(i.e. they do not even form a good first order approximation to the
(unknown) eingenbasis of the total Hamiltonian Eq. (\ref{Hamfull_j-jL-S}))
and the total \textbf{J} momentum has to be calculated as a sum of
individual \textbf{j} momenta characterizing each particle. 

We now derive the basis transformation connecting
Hamiltonian in the $\left|m_l m_s \right\rangle\,\left|m_l' m_s' \right\rangle$ independent particle basis to the Hamiltonian 
defined in the basis of the total momenta \textbf{J}. In the
\textit{jj} coupling scheme, we first use CGCs to sum up the total momenta on each site
\begin{equation}
 c_{0 \uparrow}^\dag c_{1 \uparrow}^\dag\left|0\right\rangle \rightarrow \left(\sqrt{\frac{2}{3}}\left|\frac{3}{2}\frac{1}{2}\right\rangle-\sqrt{\frac{1}{3}}\left|\frac{1}{2}\frac{1}{2}\right\rangle\right)
 \left(\left|\frac{3}{2}\frac{3}{2}\right\rangle\right),
\end{equation}
where the latter is written in the spin-orbit coupled single-particle basis $\left|j\ m_j\right\rangle$. Since we perform CG summation here independently for both electrons, we have to take Pauli 
principle into account manually by projecting out forbidden states by
hand. In the end, we arrive at:
\begin{align}
\label{U3matrix}
      & U_3 =\\& \left(\begin{smallmatrix}
 0 & 0 & 0 & 0 & 0 & 0 & \frac{\sqrt{2}}{3} & -\frac{2}{3} & 0 & -\frac{1}{3} & \frac{\sqrt{2}}{3} & 0 & 0 & 0 & 0 \\
 -\sqrt{\frac{2}{3}} & \frac{1}{\sqrt{3}} & 0 & 0 & 0 & 0 & 0 & 0 & 0 & 0 & 0 & 0 & 0 & 0 & 0 \\
 0 & 0 & -\frac{1}{\sqrt{3}} & \sqrt{\frac{2}{3}} & 0 & 0 & 0 & 0 & 0 & 0 & 0 & 0 & 0 & 0 & 0 \\
 0 & 0 & 0 & 0 & 0 & 1 & 0 & 0 & 0 & 0 & 0 & 0 & 0 & 0 & 0 \\
 0 & 0 & 0 & 0 & 0 & 0 & -\frac{1}{3} & \frac{\sqrt{2}}{3} & 0 & -\frac{\sqrt{2}}{3} & \frac{2}{3} & 0 & 0 & 0 & 0 \\
 0 & 0 & 0 & 0 & 0 & 0 & \frac{2}{3} & \frac{\sqrt{2}}{3} & 0 & -\frac{\sqrt{2}}{3} & -\frac{1}{3} & 0 & 0 & 0 & 0 \\
 0 & 0 & 0 & 0 & 0 & 0 & 0 & 0 & 0 & 0 & 0 & 0 & 1 & 0 & 0 \\
 0 & 0 & 0 & 0 & 0 & 0 & 0 & 0 & \sqrt{\frac{2}{3}} & 0 & 0 & -\frac{1}{\sqrt{3}} & 0 & 0 & 0 \\
 0 & 0 & 0 & 0 & 0 & 0 & 0 & 0 & 0 & 0 & 0 & 0 & 0 & \frac{1}{\sqrt{3}} & -\sqrt{\frac{2}{3}} \\
 \frac{1}{\sqrt{3}} & \sqrt{\frac{2}{3}} & 0 & 0 & 0 & 0 & 0 & 0 & 0 & 0 & 0 & 0 & 0 & 0 & 0 \\
 0 & 0 & \sqrt{\frac{2}{3}} & \frac{1}{\sqrt{3}} & 0 & 0 & 0 & 0 & 0 & 0 & 0 & 0 & 0 & 0 & 0 \\
 0 & 0 & 0 & 0 & 0 & 0 & \frac{\sqrt{2}}{3} & \frac{1}{3} & 0 & \frac{2}{3} & \frac{\sqrt{2}}{3} & 0 & 0 & 0 & 0 \\
 0 & 0 & 0 & 0 & 1 & 0 & 0 & 0 & 0 & 0 & 0 & 0 & 0 & 0 & 0 \\
 0 & 0 & 0 & 0 & 0 & 0 & 0 & 0 & \frac{1}{\sqrt{3}} & 0 & 0 & \sqrt{\frac{2}{3}} & 0 & 0 & 0 \\
 0 & 0 & 0 & 0 & 0 & 0 & 0 & 0 & 0 & 0 & 0 & 0 & 0 & \sqrt{\frac{2}{3}} & \frac{1}{\sqrt{3}} \\
\end{smallmatrix}
\right),\nonumber
\end{align}
which is needed for
\begin{equation}
\label{jjLSHamU3}
 \mathcal{H}^{jj}_{jj\mathrm{-basis}}=U_3^\dag \mathcal{H}^{jj}_{ls\mathrm{-basis}} U_3^{\phantom{\dagger}}
\end{equation}
to transform the Hamiltonian from the basis~(\ref{lsbasis}) into the
individual $j$ basis
$\left|j\:m_j\:j'\:m_{j'}\right\rangle$:
\begin{align}
 \hat{j}=&\big\{\left|\tfrac{1}{2}\:{-\tfrac{1}{2}}\:\tfrac{1}{2} \tfrac{1}{2} \right\rangle, \left| \tfrac{3}{2}\:\tfrac{3}{2}\:\tfrac{1}{2} \tfrac{1}{2} \right\rangle, 
 \left|\tfrac{3}{2}\:\tfrac{3}{2}\:\tfrac{1}{2} {-\tfrac{1}{2}} \right\rangle, \left| \tfrac{3}{2}\:\tfrac{1}{2}\:\tfrac{1}{2} \tfrac{1}{2} \right\rangle,\nonumber\\ 
 &
 \left|\tfrac{3}{2}\:\tfrac{1}{2}\:\tfrac{1}{2} {-\tfrac{1}{2}} \right\rangle,\left|\tfrac{3}{2}\:{-\tfrac{1}{2}}\:\tfrac{1}{2} \tfrac{1}{2} \right\rangle, \left|\tfrac{3}{2}\:{-\tfrac{1}{2}}\:\tfrac{1}{2} {-\tfrac{1}{2}} \right\rangle,
 \left|\tfrac{3}{2}\:{-\tfrac{3}{2}}\:\tfrac{1}{2}\: \tfrac{1}{2} \right\rangle,\nonumber\\  
 & \left| \tfrac{3}{2}\:{-\tfrac{3}{2}}\:\tfrac{1}{2} {-\tfrac{1}{2}} \right\rangle,
 \left|\tfrac{3}{2}\:\tfrac{1}{2}\:\tfrac{3}{2}\: \tfrac{3}{2} \right\rangle,\left|\tfrac{3}{2}\:{-\tfrac{1}{2}}\:\tfrac{3}{2} \tfrac{3}{2} \right\rangle,
 \left|\tfrac{3}{2}\:{-\tfrac{1}{2}}\:\tfrac{3}{2}\: \tfrac{1}{2} \right\rangle,\nonumber\\  
 & \left| \tfrac{3}{2}\:{-\tfrac{3}{2}}\:\tfrac{3}{2} \tfrac{3}{2} \right\rangle,
 \left|\tfrac{3}{2}\:{-\tfrac{3}{2}}\:\tfrac{3}{2}\: \tfrac{1}{2} \right\rangle, \left| \tfrac{3}{2}\:{-\tfrac{3}{2}}\:\tfrac{3}{2} {-\tfrac{1}{2}} \right\rangle\big\}^\intercal. 
\end{align}

Now, we employ the high weight decomposition method to obtain
Hamiltonian (Eq.(\ref{jjLSHamU3})) in the total $J$ basis (see Appendix
\ref{HWJJ} for details):
\begin{align}
\label{totalJfullbasisbig}
 \hat{J} = &\{S_{\phantom{1}}^{\left(\frac{1}{2},\frac{1}{2}\right)}, T_1^{\left(\frac{3}{2},\frac{1}{2}\right)}, T_0^{\left(\frac{3}{2},\frac{1}{2}\right)}, T_{-1}^{\left(\frac{3}{2},\frac{1}{2}\right)},
 M_2^{\left(\frac{3}{2},\frac{1}{2}\right)},\nonumber\\
 &M_1^{\left(\frac{3}{2},\frac{1}{2}\right)}, M_0^{\left(\frac{3}{2},\frac{1}{2}\right)}, M_{-1}^{\left(\frac{3}{2},\frac{1}{2}\right)} M_{-2}^{\left(\frac{3}{2},\frac{1}{2}\right)},
 S_{\phantom{1}}^{\left(\frac{3}{2},\frac{3}{2}\right)}, \nonumber \\
 &M_2^{\left(\frac{3}{2},\frac{3}{2}\right)},
 M_1^{\left(\frac{3}{2},\frac{3}{2}\right)}, M_0^{\left(\frac{3}{2},\frac{3}{2}\right)}, M_{-1}^{\left(\frac{3}{2},\frac{3}{2}\right)}, M_{-2}^{\left(\frac{3}{2},\frac{3}{2}\right)}\}^\intercal,
\end{align}
where $S$ is singlet state, $T_{m}$ represents a triplet state with
$J=1$, $J_z=m$, $M_{m}$ signifies a quintet state with $J=2$, $J_z=m$ and
the superscript stands for $\left(j_1,j_2\right)$.
Basis~(\ref{totalJfullbasisbig}) is equivalent to~(\ref{totalJfullbasis}) when cut down to the lowest 9 states and to~(\ref{totalJcutbasis}) upon further truncation to lowest 4 states.

In the end, we arrive at the final Hamiltonian
\begin{equation}
\label{jjfinal}
 \mathcal{H}^{jj}_{JJ\mathrm{-basis}}=U_4^\dag \mathcal{H}^{jj}_{jj\mathrm{-basis}} U_4^{\phantom{\dagger}}.
\end{equation}
where, the basis transformation is
\begin{align}
\label{U4matrix}
       U_4 =\left(\begin{smallmatrix}
 1 & 0 & 0 & 0 & 0 & 0 & 0 & 0 & 0 & 0 & 0 & 0 & 0 & 0 & 0 \\
 0 & 0 & \frac{\sqrt{3}}{2} & -\frac{1}{2} & 0 & 0 & 0 & 0 & 0 & 0 & 0 & 0 & 0 & 0 & 0 \\
 0 & 0 & 0 & 0 & \frac{1}{\sqrt{2}} & -\frac{1}{\sqrt{2}} & 0 & 0 & 0 & 0 & 0 & 0 & 0 & 0 & 0 \\
 0 & 0 & 0 & 0 & 0 & 0 & \frac{1}{2} & -\frac{\sqrt{3}}{2} & 0 & 0 & 0 & 0 & 0 & 0 & 0 \\
 0 & 1 & 0 & 0 & 0 & 0 & 0 & 0 & 0 & 0 & 0 & 0 & 0 & 0 & 0 \\
 0 & 0 & \frac{1}{2} & \frac{\sqrt{3}}{2} & 0 & 0 & 0 & 0 & 0 & 0 & 0 & 0 & 0 & 0 & 0 \\
 0 & 0 & 0 & 0 & \frac{1}{\sqrt{2}} & \frac{1}{\sqrt{2}} & 0 & 0 & 0 & 0 & 0 & 0 & 0 & 0 & 0 \\
 0 & 0 & 0 & 0 & 0 & 0 & \frac{\sqrt{3}}{2} & \frac{1}{2} & 0 & 0 & 0 & 0 & 0 & 0 & 0 \\
 0 & 0 & 0 & 0 & 0 & 0 & 0 & 0 & 1 & 0 & 0 & 0 & 0 & 0 & 0 \\
 0 & 0 & 0 & 0 & 0 & 0 & 0 & 0 & 0 & 0 & 0 & -\frac{1}{\sqrt{2}} & \frac{1}{\sqrt{2}} & 0 & 0 \\
 0 & 0 & 0 & 0 & 0 & 0 & 0 & 0 & 0 & 1 & 0 & 0 & 0 & 0 & 0 \\
 0 & 0 & 0 & 0 & 0 & 0 & 0 & 0 & 0 & 0 & 1 & 0 & 0 & 0 & 0 \\
 0 & 0 & 0 & 0 & 0 & 0 & 0 & 0 & 0 & 0 & 0 & \frac{1}{\sqrt{2}} & \frac{1}{\sqrt{2}} & 0 & 0 \\
 0 & 0 & 0 & 0 & 0 & 0 & 0 & 0 & 0 & 0 & 0 & 0 & 0 & 1 & 0 \\
 0 & 0 & 0 & 0 & 0 & 0 & 0 & 0 & 0 & 0 & 0 & 0 & 0 & 0 & 1 \\
\end{smallmatrix}
\right).
\end{align}

The correspondence between the two coupling schemes is obtained by
matrix manipulation of the above matrices $U_1$, $U_2$, $U_3$ and $U_4$,
leading to results of Eq. (\ref{renormJLScoef2}).

\section{Derivation of W-terms}
\label{AppC1}

To illustrate the renormalization of different elements of W-terms, we
consider a NNN hopping %Hamiltonian 
between the sites $i$ and $j$ which involves only hopping between
orbitals $\zeta = xy$ at each site:
\begin{equation}
\label{C10}
H =\sum_{ij\,,\sigma=\uparrow,\downarrow}t'c_{i\,\zeta \,\sigma}c_{j\, \zeta \, \sigma}. 
\end{equation}
We transform this Hamiltonian into a basis that spans the full Hilbert space of two NNN sites
$\left|l\:s\:m_l\:m_s\right\rangle_i\left|l'\:s'\:m'_l\:m'_s\right\rangle_i \otimes \left|l\:s\:m_l\:m_s\right\rangle_j + h.c.$. 
We do not explicitly show this transformed Hamiltonian $H'$ here because
of the size of the matrix ($180\times180$). The Hamiltonian in spin-orbit coupled basis within \textit{jj} coupling scheme is then calculated as
\begin{equation}
\label{C11}
H^{jj}=(U_3\times U_4)\otimes U_5)^\dag \,H'\, ((U_3\times U_4)\otimes U_5),
\end{equation}
where $U_5$ describes transformation of multiplet structure of a single
hole/electron in three $t_{2g}$ orbitals into total $j$ basis. This transformation is independent of the coupling
scheme and can be obtained easily (see, e.g.
\onlinecite{Jackeli2009,Zhong2013}).
Hamiltonian $H$ then produces another $180\times180$ matrix with quite a few non-zero entries. 
For instance, the (1,12)-th matrix element is
\begin{equation}
\label{C12}
    [H^{jj}]_{1,12}=-\sum_{\langle i,j\rangle}{\frac{t'}{6\sqrt{2}}S^{\dag}_{i}d^{\phantom{\dagger}}_{i \downarrow}d_{j \uparrow}^{\dag}T_{j 1}^{\phantom{\dagger}}},
\end{equation}
where $d_{i \sigma}^{\dag}$ stands for creating an electron on site $i$ in the $j=1/2$ doublet with $j_z=\sigma$, and $S_i$ and $T_{i\,m}$, respectively, represent the
creation of a charge excitation with singlet ($S$) and triplet ($T_m$,
$m=0,\pm1$) character on site $i$.
The resulting Hamiltonian is then mapped onto a polaronic model as
described in detail in e.g. Supplemental Material of
\onlinecite{Plotnikova2016}. 
We subsequently introduce two antiferromagnetic sublattices $A$ and $B$, and perform
the Holstein-Primakoff transformation:
\begin{equation}
\label{C13}
[H^{jj}]_{1,12}=-\sum_{\langle i,j\rangle}\frac{t'}{6\sqrt{2}}\left(S^{\dag}_{i A}T_{j 1 A}^{\phantom{\dagger}}a^{\phantom{\dagger}}_{i}+S^{\dag}_{i B}T_{j 1 B}^{\phantom{\dagger}}b^{\dagger}_{i}\right),
\end{equation}
where $a^{\dagger}_{i}(b^{\dagger}_{i})$ stands for creating a magnon on sublattice A(B).
Then, we translate it into $k$ space using Bogoluibov and Fourier
transforms and obtain
\begin{align}
\label{C14}
&[H^{jj}]_{1,12} = \nonumber \\ 
    &\frac{t'}{3} \sqrt{\frac{1}{N}} \,\Big[\, T _{1 A k} \cos \left(k_x-k_y\right) S_{A k-q}^\dag 
\left(\alpha_{-q} u_{q}+\beta_q^\dag v_{q}\right) \nonumber \\
&+ T_{1 B k} S_{B k-q}^\dag \cos \left(k_x-k_y-q_x+q_y\right)
    \left(\beta q^\dag u_{q}+\alpha_{-q} v_{q}\right) \,\Big]
\end{align}
Here, $\alpha^{\dagger}$ ($\alpha$)/$\beta^{\dagger}$ ($\beta$)
represents the magnon creation (annihiliation) operator on the two
sublattices A/B after the Bogoliubov transformation, and $u_q$ and $v_q$
are the Bogoliubov coefficients.\cite{Paerschke2017} After this transformation has been performed for all the terms of Hamiltonian, the $\frac{t'}{3} \sqrt{\frac{1}{N}}$ coefficients would enter the $W$ expressions.

In the \textit{LS} coupling scheme, the corresponding (1,12)-th element of the above hopping Hamiltonian (\ref{C10}) yields 
\begin{equation}
\label{C15}
[H^{LS}]_{1,12}=-\sum_{\langle i,j\rangle}{\frac{t'}{6\sqrt{3}}S^{\dag}_{i}d^{\phantom{\dagger}}_{i \downarrow}d_{j \uparrow}^{\dag}T_{j 1}^{\phantom{\dagger}}},
\end{equation}
rendering the renormalization of the elements of W-terms by a factor of
$\sqrt{2/3}$ compared to the hopping Hamiltonian (\ref{C12}), as discussed in Section \ref{Section:j-jL-S2sites}. 

\section{High weight decomposition method}
\label{AppB}
\subsection{\textit{LS} coupling scheme}
\label{HWLS}
We start with the ``high spin'' state with the largest possible total spin $S=1$ and highest possible \textbf{L} for this \textbf{S}.
Obviously, there are nine states with $S=1$ and $L=1$ which form the $^3P$ multiplet. From them we choose the one with the maximum 
projections $M_L$ and $M_S$: $\psi^{LS}_1=\left|S\: M_S \:L \:M_L \right\rangle = \left|1 \:1\: 1\: 1\right\rangle$. In terms 
of single-particle second quantization operators there is only one way this state can possibly be constructed:
\begin{equation}
\label{psi1LS}
 \psi^{LS}_1 = \left|1, 1, 1, 1\right\rangle = c_{0 \uparrow}^\dag c_{1 \uparrow}^\dag \left|0\right\rangle,
\end{equation}
where $c_{\alpha \sigma}^\dag$ is an operator creating an electron on the $l_{\mathrm {eff}} = 1$ orbital with $m_l=\alpha$ and spin $\sigma$, and the vacuum state $\left|0\right\rangle$ is defined as
empty $t_\mathrm{2g}$ shell.
To construct the next possible state we employ a ladder operator $\hat{L}^{-}$:
\begin{equation}
\label{psi2ls}
 \psi^{LS}_2 = \hat{L}^{-} \psi^{LS}_1.
\end{equation}
Using formula for the ladder operator known from textbooks (see for example Landau and Lifshitz~\cite{LandauLifshitz}) 
\begin{equation}
    \left\langle L, M_{L-1} \right|\hat{L}^{-} \left|L, M_L \right\rangle = \sqrt{(L+M_L)(L-M_L+1)},
\end{equation}
and normalizing~(\ref{psi2ls}) we get
\begin{equation}
  \psi^{LS}_2=\left|1\:1\:1\:0\right\rangle = c_{-1 \uparrow}^\dag c_{1 \uparrow}^\dag \left|0\right\rangle.
\end{equation}

Now we can either apply $\hat{L}^{-}$ once more or employ spin ladder $\hat{S}^{-}$ operator instead. Let us look at the effect of the latter:
\begin{equation}
  \left|1\:0\:1\:0 \right\rangle=\hat{S}^{-} \psi^{LS}_2= \frac{1}{\sqrt{2}}\left(c_{-1 \downarrow}^\dag c_{1 \uparrow}^\dag + c_{-1 \uparrow}^\dag c_{1 \downarrow}^\dag \right)\left|0\right\rangle.
\end{equation}

For a particular electronic configuration containing indistinguishable electrons according to the empirical Hund's rule the ground state is the one
with the largest possible for this configuration value of the total spin $S$ and the largest possible for this $S$ value of the total orbital momentum $L$.
So, having obtained all nine states of the $^3P$ multiplet in this way we proceed by searching for a state with the highest possible total orbital momentum. 
Since one has to place two electrons on the same orbital to get total orbital momentum $L=2$, they must have opposite spins in order to obey Pauli's principle.
This state thus has $L=2$, $S=0$ and belongs to $^1D$ quintet. Again, for the state with the highest possible momentum, be it orbital or spin, there is always one unique way to construct it:
\begin{equation}
 \psi^{LS}_{10}=\left|0\:0\:2\:2\right\rangle = c_{1 \downarrow}^\dag c_{1 \uparrow}^\dag \left|0\right\rangle.
\end{equation}

It is important on this step to keep operator ordering convention consistent with that used in~(\ref{psi1LS}). After we have obtained all five states of $^1D$ multiplet using ladder operators, we only need to find
the last missing state: singlet $^1S$ (full list of multiplets forming
for a particular electronic configuration can be found in many atomic
physics book, see e.g. Table 2.1 in [\onlinecite{Sobelman}]).
We know that $^1S$ state shall have $M_S=0$ and $M_L=0$, but we do not know what the quantum numbers $L$, $S$ are. What we however know is that $^1S$ state has to be orthogonal to the other two
states with $M_S=0$ and $M_L=0$, which are written as
\begin{align}
\label{twosingletsLS}
 \left|1\:0\:1\:0 \right\rangle & = \frac{1}{\sqrt{2}}\left(c_{-1 \downarrow}^\dag c_{1 \uparrow}^\dag+c_{-1 \uparrow}^\dag c_{1 \downarrow}^\dag\right)\left|0\right\rangle, \\
 \left|0\:0\:2\:0 \right\rangle & = \frac{1}{\sqrt{6}}\left(c_{-1 \downarrow}^\dag c_{1 \uparrow}^\dag-c_{-1 \uparrow}^\dag c_{1 \downarrow}^\dag+
 2c_{0 \downarrow}^\dag c_{0 \uparrow}^\dag\right)\left|0\right\rangle.\nonumber
\end{align}

Since there can be no other combination of two creation operators creating a state with both $M_S=0$ and $M_L=0$ other than the three used in~(\ref{twosingletsLS}) the missing state has to be a combination of
them as well and simultaneously orthogonal to the two states in~(\ref{twosingletsLS}). Employing trivial linear algebra we get that the $^1S$ multiplet is written as
\begin{equation}
 \psi^{LS}_{15}=\left|0\:0\:0\:0\right\rangle =  \frac{1}{\sqrt{3}}\left(c_{-1 \downarrow}^\dag c_{1 \uparrow}^\dag-c_{-1 \uparrow}^\dag c_{1 \downarrow}^\dag-
 c_{0 \downarrow}^\dag c_{0 \uparrow}^\dag\right)\left|0\right\rangle.
\end{equation}

\subsection{\textit{jj} coupling scheme}
\label{HWJJ}

We start from the state with highest possible  $M_J=2$. The state with the highest total momenta $J=2$ can be constructed either by placing one electron on the $j=\frac{3}{2}$ state 
with energy $\lambda=\xi/2$ and one electron on the $\frac{1}{2}$ state or by placing two electrons on $j=\frac{3}{2}$ quartets both having energy $\lambda=\xi/2$
so that a two-particle state has energy $\lambda=\xi$. Let us
start with a state that is lower in energy
\begin{equation}
   \psi^{jj}_5 = \left|J=2\:M_J=2 \right\rangle^{\left(\frac{3}{2},\frac{1}{2}\right)} =  \left| \frac{3}{2}\:\frac{3}{2}\:\frac{1}{2}\: \frac{1}{2} \right\rangle.
\end{equation}

Applying ladder operator $J^{-}$ and normalizing the result we obtain the next state
\begin{align}
    &\psi^{jj}_6 = \left|J=2\:M_J=1 \right\rangle^{\left(\frac{3}{2},\frac{1}{2}\right)} =  \\
   &\frac{\sqrt{3}}{2}\left| \frac{3}{2}\:\frac{1}{2}\:\frac{1}{2}\:\frac{1}{2} \right\rangle+\frac{1}{2}\left| \frac{3}{2}\:\frac{3}{2}\:\frac{1}{2}\: {-\frac{1}{2}} \right\rangle.
\end{align}

Once we have obtained five possible $J=2$ states we consider the other $\left|J=2\:M_J=1 \right\rangle$ configuration formed by two electrons in the $j=\frac{3}{2}$ quartet:
\begin{equation}
  \psi^{jj}_{11} = \left|J=2\:M_J=2 \right\rangle^{\left(\frac{3}{2},\frac{3}{2}\right)} =  \left| \frac{3}{2}\:\frac{1}{2}\:\frac{3}{2}\:\frac{1}{2} \right\rangle.
\end{equation}
Note that once chosen, the ordering convention has to be followed since fermionic operators anticommute. 

Rest of the derivation is performed analogously to that in section
\ref{HWLS}.

\section{$t-J$ model within the {\it LS} coupling scheme}
\label{AppC}

The kinetic part of the $t-J$ in the $LS$ coupling scheme is
\begin{align}
	    \label{Hparts}
	    &{\mathcal{H}}^{\bf {\rm \bf  h}}_{t}\!= \! \sum\limits_{{\bf k}} \left( {\bf h}_{{\bf k} \rm A}^{\dagger}\hat{V}^{0}_{{\bf k}} {\bf h}_{{\bf k} \rm A}\! +\!{\bf h}_{{\bf k} \rm B}^{\dagger}\hat{V}^{0}_{\bf k} {\bf h}_{{\bf k} \rm B} \right)\!\\
	    &+\! \sum\limits_{{\bf k}, {\bf q}} \left(  {\bf h}_{{\bf k-q} \rm B}^{\dagger} \hat{V}^{\alpha}_{{\bf k},{\bf q}} {\bf h}_{{\bf k} \rm B} \alpha_{\bf q}^{\dagger}  \!+\!
  {\bf h}_{{\bf k-q} \rm A}^{\dagger}  \hat{V}^{\beta}_{{\bf k},{\bf q}} {\bf h}_{{\bf k} \rm B} \beta_{\bf q}^{\dagger} \!+\! h.c. \right)\!.\nonumber
   \end{align}  
where the free hopping matrix is defined as
\begin{align}
     \label{V0}
      &\hat{V}^{0}_{\bf k}= \begin{pmatrix}
      F_1 & 0 & -F_2 & 0 &   0 & P_2 & 0 & -P_1\\
      0 & F_4 & 0 & 0 &      P_1 & 0 & Q_1 & 0 \\
      -F_2 & 0 & F_3 & 0 &   0 & Q_2 & 0 & Q_1 \\
      0 & 0 & 0 & 0 &        -P_2 & 0 & Q_2 & 0 \\
      0 & P_1 & 0 & -P_2 &   F_1 & 0 & F_2 & 0\\
      P_2 & 0 & Q_2 & 0  &   0 & 0 & 0 & 0  \\
      0 & Q_1 & 0 & Q_2 &    F_2 & 0 & F_3 & 0   \\
      -P_1 & 0 & Q_1 & 0 &   0 & 0 & 0 & F_4  \\
      \end{pmatrix},
 \end{align}
and the the matrices containing vertices are
  \begin{align}
       \label{Va}
      &\hat{V}^{\alpha}_{{\bf k},{\bf q}}= \begin{pmatrix}
      0 & L_3 & 0 & -L_3 &  Y_1 & 0 & -W_2 & 0\\
      L_3 & 0 & L_1 & 0 &  0 & Y_4 & 0 & W_1 \\
      0 & L_1 & 0 & L_1 &  -W_2 & 0 & Y_2 & 0 \\
      -L_3 & 0 & L_1 & 0 &  0 & W_1 & 0 & Y_3 \\
      0 & 0 & 0 & 0   & 0 & L_4 & 0 & -L_4 \\
      0 & 0 & 0 & 0   & L_4 & 0 & L_2 & 0  \\
      0 & 0 & 0 & 0   & 0 & L_2 & 0 & L_2 \\
      0 & 0 & 0 & 0   & -L_4 & 0 & L_2 & 0  \\
      \end{pmatrix}, 
 \end{align}  
  \begin{align}
       \label{Vb}
      &\hat{V}^{\beta}_{{\bf k},{\bf q}}= \begin{pmatrix}
      0 & L_4 & 0 & -L_4 &  0 & 0 & 0 & 0 \\
      L_4 & 0 & L_2 & 0 &  0 & 0 & 0 & 0 \\
      0 & L_2 & 0 & L_2 &  0 & 0 & 0 & 0 \\
      -L_4 & 0 & L_2 & 0 &  0 & 0 & 0 & 0 \\
     Y_1 & 0 & W_2 & 0   & 0 & L_3 & 0 & -L_3  \\
     0 & Y_3 & 0 & W_1   & L_3 & 0 & L_1 & 0  \\
     W_2 & 0 & Y_2 & 0   & 0 & L_1 & 0 & L_1  \\
     0 & W_1 & 0 & Y_4   & -L_3 & 0 & L_1 & 0  \\
      \end{pmatrix},
       \end{align}

\section{Free hopping and vertex elements}
\label{AppD}

The nearest neighbor free hopping $P({\bf k})$, $Q({\bf k})$ and the polaronic diagonal $Y({\bf k},{\bf q})$ and non-diagonal $W({\bf k},{\bf q})$ vertex elements are
     \begin{flalign} 
   &P_1({\bf k})=\frac{2\left(2t_1-t_2\right)}{3\sqrt{3}}\gamma_{\bf k}-\frac{2t_3}{3\sqrt{3}}\gamma_{\bf k}, \\
   &P_2({\bf k})=\frac{2t_2}{\sqrt{3}}\tilde{\gamma}_{\bf k}-\frac{2t_3}{\sqrt{3}}\tilde{\gamma}_{\bf k},\\
   &Q_1({\bf k})=\frac{\left(4t_1+t_2\right)}{3\sqrt{2}}\gamma_{\bf k}+\frac{t_3}{3\sqrt{2}}\gamma_{\bf k},\\
      &Q_2({\bf k})=\frac{t_2}{\sqrt{2}}\tilde{\gamma}_{\bf k}-\frac{t_3}{\sqrt{2}}\tilde{\gamma}_{\bf k},\\
   &W_1({\bf k},{\bf q})=\frac{t_3-t_2}{\sqrt{2N}}\left(\tilde{\gamma}_{\bf k-q}u_{\bf q}+\tilde{\gamma}_{{\bf k}}v_{\bf q}\right),\\
   &W_2({\bf k},{\bf q})=-\frac{4\left(2t_1-t_2-t_3\right)}{3\sqrt{3N}}\left(\gamma_{{\bf k-q}}u_{\bf q}-\gamma_{{\bf k}}v_{\bf q}\right),\\
   &Y_1({\bf k},{\bf q})=-\frac{16\left(t_1+t_2+t_3\right)}{9\sqrt{2N}}\left(\gamma_{{\bf k-q}}u_{\bf q}+\gamma_{{\bf k}}v_{\bf q}\right),\\
   &Y_2({\bf k},{\bf q})=-\frac{2\left(4t_1+t_2+t_3\right)}{3\sqrt{2N}}\left(\gamma_{{\bf k-q}}u_{\bf q}+\gamma_{{\bf k}}v_{\bf q}\right),\\
   &Y_3({\bf k},{\bf q})=-\frac{4t_1+t_2+t_3}{3\sqrt{2N}}\gamma_{{\bf k}}v_{\bf q}-\frac{3\left(t_2+t_3\right)}{\sqrt{2N}}\gamma_{{\bf k-q}}u_{\bf q},\\
   &Y_4({\bf k},{\bf q})=-\frac{4t_1+t_2+t_3}{3\sqrt{2N}}\gamma_{{\bf k-q}}u_{\bf q}-\frac{3\left(t_2+t_3\right)}{\sqrt{2N}}\gamma_{{\bf k}}v_{\bf q},
   \end{flalign}  
   with  $\gamma_{{\bf k}}=1/2(\cos{k_x}+\cos{k_y})$ and
   $\tilde{\gamma}_{{\bf k}}=1/2(\cos{k_x}-\cos{k_y})$, where $N$ is the number of sites, and
   $u_{\bf q}$ and $v_{\bf q}$ are the Bogoliubov coefficients.\cite{Paerschke2017}
   
The free hopping elements arising from the next-nearest and third neighbor hoppings are
     \begin{flalign} 
   &F_1({\bf k})=-\frac{4 t'\gamma'_{\bf k}}{9}-\frac{4 t''\gamma''_{\bf k}}{9},\\
   &F_2({\bf k})=-\frac{8 t'\gamma'_{\bf k}}{3\sqrt{6}}-\frac{8 t''\gamma''_{\bf k}}{3\sqrt{6}},\\
   &F_3({\bf k})=-\frac{2t'\gamma'_{\bf k}}{3}-\frac{2t''\gamma''_{\bf k}}{3},\\
   &F_4({\bf k})=-\frac{t'\gamma'_{\bf k}}{3}-\frac{t''\gamma''_{\bf k}}{3},
    \end{flalign}  
where $\gamma'_{{\bf k}}=\cos{k_x}\cos{k_y}$.
The polaronic next-nearest and third neighbor vertex elements are 
     \begin{flalign} 
   &L_1({\bf k},{\bf q})=\frac{4 t'}{3\sqrt{N}}\gamma'_{{\bf k-q}}u_{\bf q}+\frac{4 t''}{3\sqrt{N}}\gamma''_{{\bf k-q}}u_{\bf q},\\
   &L_2({\bf k},{\bf q})=\frac{4 t'}{3\sqrt{N}}\gamma'_{{\bf k}}v_{\bf q}+\frac{4 t''}{3\sqrt{N}}\gamma''_{{\bf k}}v_{\bf q},\\
   &L_3({\bf k},{\bf q})=\frac{8 t'}{3\sqrt{6N}}\gamma'_{{\bf k-q}}u_{\bf q}+\frac{8 t''}{3\sqrt{6N}}\gamma''_{{\bf k-q}}u_{\bf q},\\
   &L_4({\bf k},{\bf q})=\frac{8 t'}{3\sqrt{6N}}\gamma'_{{\bf k}}v_{\bf q}+\frac{8 t''}{3\sqrt{6N}}\gamma''_{{\bf k}}v_{\bf q}
    \end{flalign}  
with $\gamma''_{{\bf k}}=1/2(\cos{2 k_x}+\cos{2 k_y})$.

%\bibliography{sio_arpes}
\input{revised_ms.bbl}

\end{document}